\documentclass[authoryear,final,3p,times,twocolumn]{elsarticle}

\usepackage{epsfig}

\usepackage{natbib}

\usepackage{amssymb}

\usepackage{lineno}
\usepackage{rotating}

\newcommand{\beq}{\bigskip\begin{equation}}
\newcommand{\eeq}{\bigskip\end{equation}}

\newcommand{\m}{$\mu$m}
\newcommand{\pderiv}[2]{\frac{\partial #1}{\partial #2}}

\journal{Icarus}

\begin{document}

\begin{frontmatter}


\title{Neptune at Summer Solstice: Zonal Mean Temperatures from Ground-Based Observations, 2003-2007}

\author[ox]{Leigh N. Fletcher}
\ead{fletcher@atm.ox.ac.uk}
\author[idp]{Imke de Pater}
\author[jpl]{Glenn S. Orton}
\author[hbh]{Heidi B. Hammel}
\author[sit]{Michael L. Sitko}
\author[ox]{Patrick G.J. Irwin}





\address[ox]{Atmospheric, Oceanic \& Planetary Physics, Department of Physics, University of Oxford, Clarendon Laboratory, Parks Road, Oxford, OX1 3PU, UK}
\address[idp]{Astronomy Department, 601 Campbell Hall, University of California, Berkeley, CA 
94720, USA}
\address[hbh]{ Association of Universities for Research in Astronomy, 1212 New York Avenue NW, Suite 450, Washington, DC 20005, USA; Space Science Institute, 4750 Walnut Street, Suite 205, Boulder, CO 80301, USA}
\address[sit]{Department of Physics, University of Cincinnati, Cincinnati, OH 45221, USA}
\address[jpl]{Jet Propulsion Laboratory, California Institute of Technology, 4800 Oak Grove Drive, Pasadena, CA, 91109, USA}




\begin{abstract}
Imaging and spectroscopy of Neptune's thermal infrared emission from Keck/LWS (2003), Gemini-N/MICHELLE (2005); VLT/VISIR (2006) and Gemini-S/TReCS (2007) is used to assess seasonal changes in Neptune's zonal mean temperatures between Voyager-2 observations (1989, heliocentric longitude $L_s=236^\circ$) and southern summer solstice (2005, $L_s=270^\circ$).  Our aim was to analyse imaging and spectroscopy from multiple different sources using a single self-consistent radiative-transfer model to assess the magnitude of seasonal variability.  Globally-averaged stratospheric temperatures measured from methane emission tend towards a quasi-isothermal structure (158-164 K) above the 0.1-mbar level, and are found to be consistent with spacecraft observations of AKARI.  This remarkable consistency, despite very different observing conditions, suggests that stratospheric temporal variability, if present, is $<\pm5$ K at 1 mbar and $<\pm3$ K at 0.1 mbar during this solstice period. Conversely, ethane emission is highly variable, with abundance determinations varying by more than a factor of two (from 500 to 1200 ppb at 1 mbar).  The retrieved C$_2$H$_6$ abundances are extremely sensitive to the details of the $T(p)$ derivation, although the underlying cause of the variable ethane emission remains unidentified.  Stratospheric temperatures and ethane are found to be latitudinally uniform away from the south pole (assuming a latitudinally-uniform distribution of stratospheric methane), with no large seasonal hemispheric asymmetries evident at solstice.  At low and mid-latitudes, comparisons of synthetic Voyager-era images with solstice-era observations suggest that tropospheric zonal temperatures are unchanged since the Voyager 2 encounter, with cool mid-latitudes and a warm equator and pole.  A re-analysis of Voyager/IRIS 25-50 $\mu$m mapping of tropospheric temperatures and para-hydrogen disequilibrium (a tracer for vertical motions) suggests a symmetric meridional circulation with cold air rising at mid-latitudes (sub-equilibrium para-H$_2$ conditions) and warm air sinking at the equator and poles (super-equilibrium para-H$_2$ conditions).   The most significant atmospheric changes have occurred at high southern latitudes, where zonal temperatures retrieved from 2003 images suggest a polar enhancement of 7-8 K above the tropopause, and an increase of 5-6 K throughout the $70-90^\circ$S region between 0.1 and 200 mbar. Such a large perturbation, if present in 1989, would have been detectable by Voyager/IRIS in a single scan despite its long-wavelength sensitivity, and we conclude that Neptune's south polar cyclonic vortex increased in strength significantly from Voyager to solstice.

\end{abstract}


\end{frontmatter}


\section{Introduction}
\label{intro}

As the furthest planet from our Sun, Neptune provides an extreme test of our understanding of seasonal processes and atmospheric photochemistry.  Neptune's complex meteorology is driven by a balance between its intrinsic luminosity and absorption of sunlight by methane and aerosols in the upper troposphere.    Visible and near-infrared imaging of Neptune from Voyager \citep{89smith, 11karkoschka}, the Hubble Space Telescope \citep{95sromovsky, 95hammel, 01sromovsky, 11karkoschka_ch4}, and ground-based observatories \citep{98roddier, 03max, 02gibbard, 03gibbard, 10statia, 11irwin}, have shown the planet to be dynamically active despite its large distance from the Sun.  Unlike Uranus, with its unusual inclination and negligible internal heat source \citep[e.g.,][]{91pearl}, Neptune's weather layer exhibits rapidly varying cloud activity, zonal banding, dark ovals and sporadic orographic clouds.    Furthermore, the extent of this meteorological activity and the planet's global visible albedo appear to vary with time \citep{06lockwood, 07hammel_var}.  In this study we attempt to connect this cloud-level activity to changes in Neptune's thermal structure and atmospheric chemistry between Voyager observations (1989) and ground-based observations close to Neptune's 2005 summer solstice.

Spectroscopy and imaging in the thermal infrared is challenging due to Neptune's cold temperatures and small angular size as viewed from Earth.  Nevertheless this provides our only method of deducing the atmospheric thermal structure and composition above the cloud tops.  Mid-infrared wavelengths are dominated by the collision-induced opacity of H$_2$ and He, along with the collection of hydrocarbons resulting from a chain of photochemical reactions initiated by UV photolysis of methane \citep[e.g.,][]{87orton, 05moses_jup}.  Ethane at 12 $\mu$m was first observed by \citet{77gillett}, and together with methane at 7.7 $\mu$m dominates the N-band (7-13 $\mu$m) spectrum of Neptune \citep{92orton}.  Photometric imaging or spectroscopy is sensitive to emission from the stratosphere (0.1 mbar) down to the upper troposphere (200 mbar), and the measurement of a broad thermal-IR wavelength range allows us to disentangle the various contributions (temperatures, composition, aerosols) to the upwelling radiance.  The latitudinal variability of Neptune's troposphere was first derived from Voyager/IRIS observations in 1989, showing a warm equator, cool mid-latitudes and a moderate increase in emission towards the south pole \citep{89conrath, 91bezard, 91conrath}.  This pattern of emission qualitatively supports a residual-mean circulation of upwelling and adiabatic cooling at mid-latitudes (consistent with near-IR imaging showing cloud activity concentrated at southern mid-latitudes), with subsidence at the equator and south pole \citep{91conrath, 98conrath}.  

Voyager/IRIS lacked the mid-IR sensitivity to study Neptune's hydrocarbon emission, which was later used by space-based observatories to determine the disc-averaged stratospheric properties from the Infrared Space Observatory \citep[ISO, ][]{99bezard, 99schulz, 03fouchet}, Spitzer Space Telescope \citep{08meadows} and ISAS/JAXA's AKARI infrared astronomy satellite \citep{10fletcher_akari}.  \citet{03burgdorf} combined Neptune's far-IR spectrum from ISO (28-145 $\mu$m) with ground-based observations between 17-24 $\mu$m \citep{87orton, 90orton} to study Neptune's disc-averaged tropospheric temperatures.  Significantly larger primary mirrors are required to move beyond the disc-average to spatially-resolved imaging, such that the latitudinal and temporal variation of Neptune's tropospheric and stratospheric temperatures can be observed.  In this study we attempt a synthesis of ground-based mid-IR observations from 8-10-m class telescopes over a five-year period surrounding Neptune's southern summer solstice (2005).  This will be compared to new atmospheric models for Voyager/IRIS results in 1989 and AKARI results in 2007.  The results will be used to search for spatial and temporal variations in Neptune's temperatures and composition between 1989 and 2007.  

This paper is organised into three broad sections, starting with the development of self-consistent stratospheric temperature profiles and ethane abundances (both disc-integrated and latitudinally-resolved) for Keck, AKARI and Gemini observations of Neptune's 7-13 $\mu$m emission between 2003 and 2007 (Section \ref{stratos}).  Stratospheric temperatures are then incorporated into new models for Neptune's tropospheric temperature and para-hydrogen structure from fitting Voyager/IRIS data from 1989 (Section \ref{tropos}), and comparing this with new ground-based observations of Neptune's 17-25 $\mu$m emission in 2003.  Finally, the meridional temperature structure is used to generate synthetic images in Section \ref{images} to compare with photometric imaging from 2003 to 2007 from Keck \citep{13depater}, Gemini-North \citep{07hammel}, VLT \citep{07orton} and Gemini-South \citep{12orton} to assess thermal variability since the Voyager encounters.  A  solstice-era latitudinal temperature structure will be derived from these ground-based images, and the end product will be a consistent survey of Neptunian temperatures from the time of Voyager to the epoch surrounding the summer solstice.


\section{Radiative Transfer Modelling}
\label{model}

A flexible radiative-transfer and retrieval algorithm is required to analyse multiple sources of thermal-IR Neptune observations using a single, consistent model.  The NEMESIS software suite \citep{08irwin} offers an optimal estimation retrieval architecture \citep{00rodgers} for the inversion of remote sensing data to deduce atmospheric profiles, whilst allowing the user to pre-compute ranked tables of absorption coefficients \citep[$k$-distributions, ][]{91lacis} in a manner specific to individual instruments.  In this study, we pre-calculate $k$-distributions for all of the infrared imaging filters used on Keck, Gemini and VLT (see Section \ref{images}), and for the spectral resolutions provided by the Voyager, Keck and Gemini spectroscopy.  Despite instrumental differences, the \textit{a priori} atmospheric state and the optimal estimation technique remain identical, allowing a robust comparison between rather different data sources.  Furthermore, by using the correlated-$k$ method for the calculation of spectra, NEMESIS allows rapid inversion of observations to converge on a family of atmospheric models (temperature and composition) consistent with the available data.  It should be noted, however, that inversion of thermal-IR spectra remains an inherently degenerate problem (as both atmospheric temperatures and molecular abundances contribute to the emission spectrum), and these caveats will be discussed in the following sections.

The \textit{a priori} atmospheric state provides a starting point for the retrieval process, preventing over-fitting and the occurrence of non-physical oscillations in the vertical atmospheric profiles.  Here we utilise the stratospheric $T(p)$ profile from analysis of AKARI observations of Neptune \citep{10fletcher_akari} as a starting point, with tropospheric temperatures based on a combination \citep{05moses_jup} of Voyager radio science observations \citep{92lindal_nep} and ground-based 17-24 $\mu$m spectroscopy in the 1980s \citep{87orton, 90orton}.  Temperature and the fraction of ortho-to-para hydrogen were defined on 100 levels, equally spaced in $\log(p)$.  \citet{10fletcher_akari} demonstrated the extreme degeneracy between the stratospheric $T(p)$ and the assumed CH$_4$ mole fraction above the tropopause \citep[$(0.9\pm0.3)\times10^{-3}$ at 50 mbar,][]{10fletcher_akari}, which was poorly known at the time.  Subsequent analysis of Herschel sub-millimetre observations of methane rotational lines \citep{10lellouch} found a best-fitting CH$_4$ abundance of $1.5\times10^{-3}$.  Following \citet{11greathouse}, we adopt the Herschel-derived CH$_4$ abundance for the stratosphere.  The deep mole fraction was set globally to 2.2\% \citep[determined from a combination of Voyager/RSS radio occultations and ground-based measurements of the H$_2$ quadrupole lines,][]{94baines}, decreasing towards the tropopause following a saturation with 100\% relative humidity, and then rising above the tropopause to a stratospheric value of $1.5\times10^{-3}$ at 40 mbar.  However, note that the stratospheric spectra considered here are insensitive to the choice of tropospheric methane humidity and the deep mole fraction, which could rise to as high as 4.0\% at low latitudes \citep[equivalent to a globally-averaged value of 3.4\%,][]{11karkoschka_ch4}.  At the highest altitudes, the methane abundance decreases due to both diffusive processes and photochemical destruction towards the homopause, as reviewed by \citet{05moses_jup}.  The stratospheric methane abundances are assumed to be latitudinally uniform, as mid-IR spectroscopy is largely insensitive to the tropospheric methane variations described by \citet{11karkoschka_ch4}.  However, stratospheric temperatures are extremely sensitive to this assumption, and would differ depending on equatorial injections of methane \citep[e.g.,][]{11karkoschka_ch4,11greathouse} or polar injections \citep{07orton}.  To ease comparison with independent analysis of 2007 Gemini data by \citet{11greathouse}, we choose to scale the hydrocarbon distributions of \citet{05moses_jup} 'Model C' photochemical model of Neptune's stratosphere, specifically focussing on the vertical distribution of ethane.  We hold the $^{12}$C/$^{13}$C ratio constant at the terrestrial value of 89, and the D/H ratio in methane constant at $(3.0\pm1.0)\times10^{-4}$ from AKARI analysis \citep{10fletcher_akari}, consistent with the $2.64^{+0.64}_{-0.56}\times10^{-4}$ abundance inferred from Herschel HD measurements \citep{13feuchtgruber}.  The He/H$_2$ mixing ratio was held at 0.18 for all levels \citep[i.e., equivalent to a helium mole fraction of 0.15 and a hydrogen mole fraction of 0.83 when 2.2\% CH$_4$ is assumed in the deep atmosphere,][]{93conrath_nep}. Unless stated otherwise, we consider the ratio of ortho-to-para hydrogen to be in thermal equilibrium.

The collision-induced continuum shaping Neptune's thermal-IR spectrum used H$_2$-H$_2$ opacities from \citet{07orton}, who updated the calculations of \citet{85borysow} for the low temperatures of the ice giants, plus additional H$_2$-He, H$_2$-CH$_4$ and CH$_4$-CH$_4$ opacities from \citet{88borysow, 86borysow, 87borysow}.  Furthermore, M. Gustafsson (\textit{pers. comms.}) provided \textit{ab initio} calculations of absorption coefficients for a range of para-hydrogen fractions appropriate to the ice giants, which were sampled on a $\Delta f_p$ grid of 0.02 to allow accurate reproduction of the collision-induced continuum.  Line data for methane and its isotopologues was taken from \citet{03brown}, H$_2$ broadened using a half width of 0.059 cm$^{-1}$atm$^{-1}$ at 296 K and a temperature dependence $T^n$ where $n=0.44$ \citep{93margolis}.  Ethane line parameters were provided by \citet{07vander}, with a half width of 0.11 cm$^{-1}$atm$^{-1}$ at 296 K \citep{87blass} and $n=0.94$ \citep{88halsey}.  All other gases (including C$_2$H$_2$ and C$_2$H$_4$) were extracted from the GEISA 2003 database \citep{05geisa}.  These data were used to generate $k$-distributions from line-by-line spectra (ranking absorption coefficients according to their frequency distributions within a defined interval) using the direct sorting method of \citet{89goody_ck} and instrument functions relevant to each observation (i.e., uniform grids for spectroscopy, filter passbands for imaging).

\section{Stratospheric N-Band Spectroscopy}
\label{stratos}

\begin{figure*}[tbp]
\centering
\includegraphics[width=15cm]{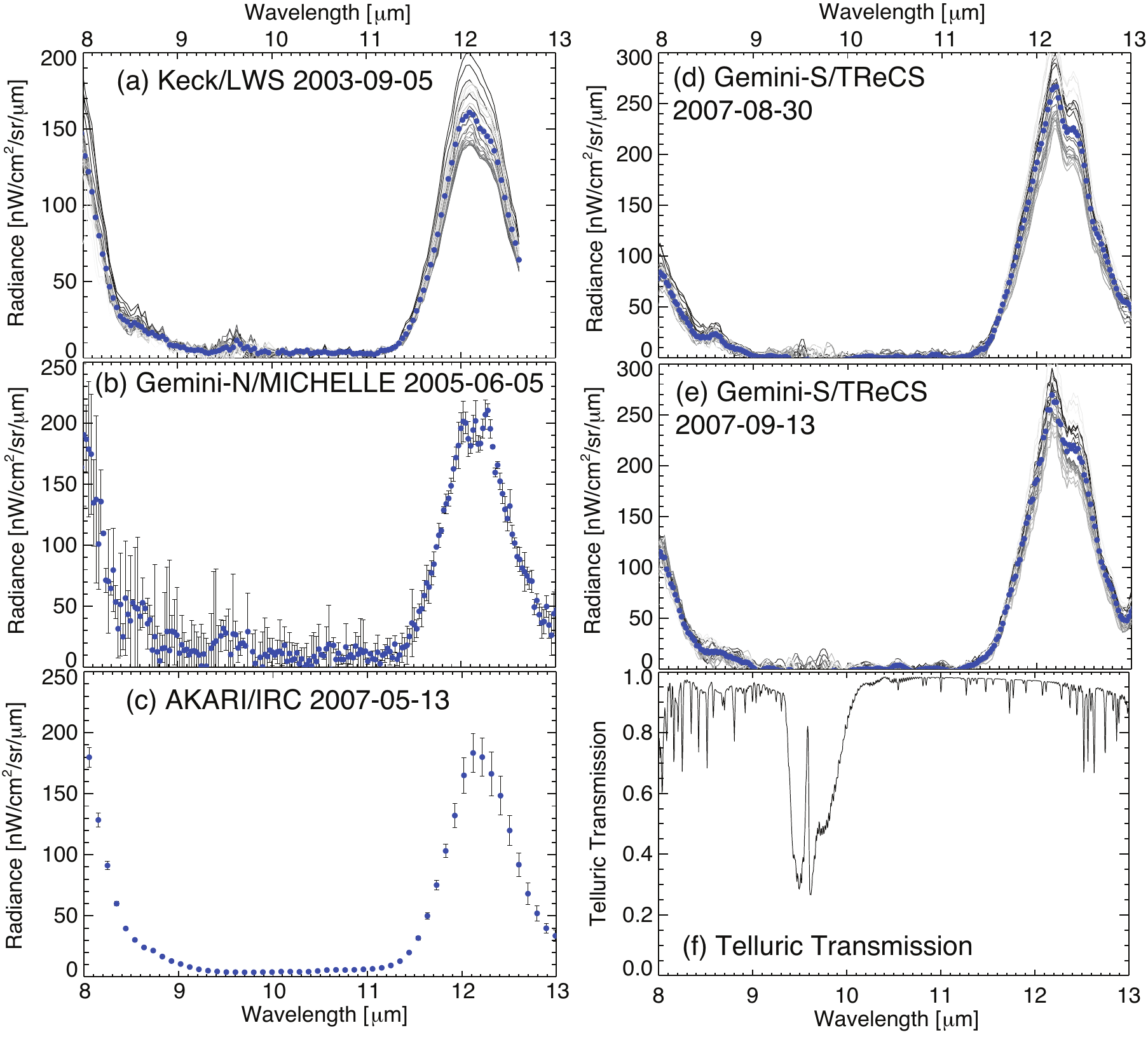}
\caption{The five examples of low-resolution ($R\approx100-200$) N-band spectroscopy considered in this study.  Panels (a), (d) and (e) were extracted from spatially-resolved spectroscopy with the slit aligned along the central meridian; grey lines show the spectrum for each pixel from south pole (dark) to north pole (light); the circles are a mean over all visible latitudes.  Panels (b) and (c) are disc-integrated spectra from AKARI/IRC and Gemini-N/MICHELLE, where the vertical bars are estimates of the uncertainty on each datapoint (circles).  This figure highlights the variable size of the methane (7.7 $\mu$m) and ethane (12.3 $\mu$m) bands with time.  Panel (f) shows a typical Mauna Kea transmission spectrum for comparison \citep{92lord}.  Noisy data between 9-11 $\mu$m were omitted from this study, particularly the region of telluric ozone absorption near 9.6 $\mu$m. }
\label{meanspx}
\end{figure*}

Our aim in this section is to determine the meridional variability of Neptune's stratospheric temperatures and ethane abundances from low spectral resolution ($R\approx100$) N-band 7-13 $\mu$m observations from Keck (2003) and Gemini (2007).  This will be compared to disc-averaged Gemini (2005) and AKARI (2007) observations, and to high spectral resolution Gemini/TEXES observations from October 2007 reported elsewhere \citep{11greathouse} to develop a consistent structure for Neptune's stratosphere as a prerequisite for tropospheric modelling (Section \ref{tropos}) and synthetic image generation (Section \ref{images}). All five sources of N-band spectroscopy are shown in Fig. \ref{meanspx}.  

\subsection{Sources of 7-13 $\mu$m spectroscopy}

\textbf{AKARI (2007): } Neptune observations between 7.4-13.0 $\mu$m recorded by ISAS/JAXA's AKARI infrared astronomy satellite \citep{07onaka} are briefly revisited to provide a disc-averaged $T(p)$ structure consistent with the latest Herschel determinations of stratospheric methane.  Full details of the AKARI spectrum were presented by \citet{10fletcher_akari}.  We utilise spectra obtained by the SG2 (7.4-13 $\mu$m) grism module of the IRC (Infrared Camera) with a resolution of $R=34$ at 10.6 $\mu$m on May 13, 2007.  Unlike terrestrial spectra (where telluric absorption obscures the 7.4-8.0 $\mu$m spectrum from view), the AKARI data provide access to the full CH$_4$ band at 7.7 $\mu$m, albeit in a disc-averaged sense \citep{10fletcher_akari}.  

\textbf{Keck (2003): } The combination of the 10-m W.M. Keck I observatory and the now-decommissioned Long Wave Spectrometer (LWS) provided the highest mid-infrared spatial resolutions on Neptune ever obtained from Earth.  LWS spectroscopy between 8.1-12.9 $\mu$m was recorded on September 6, 2003, two years before the summer solstice when Neptune subtended 2.3".  Full details of the data acquisition and reduction are provided by \citet{13depater}.  The 10.24" slit was sufficient to measure the zonal emission profile in a single observation (06:15-06:50 UT, 576 seconds of integration) and a second observation measured the east-west centre-to-limb profile (07:05-07:43 UT).  The slit width was 6 pixels, with the 0.0847 arcsec/pixel plate scale this corresponds to a width of 0.5".  A total of 128 pixels on the Boeing Si:As array covers the wavelength range, providing a resolution of $R\approx100$ in the N-band.   The standard stars HD198542 and HD188154 \citep{99cohen} were used for both telluric corrections and photometric calibration, achieving a photometric accuracy of 10-15\%.

\textbf{Gemini-N (2005):} On July 4 and 5, 2005, close to Neptune's summer solstice, the MICHELLE mid-infrared imager and spectrometer \citep{97glasse} on the Gemini-North 8.1-m telescope obtained low-resolution spectroscopy of Neptune's N-band emission between 7-14 $\mu$m with a spectral resolution of $R=200$ (0.05 $\mu$m at 10 $\mu$m).  The observations used a plate scale of 0.183"/pixel, a slit length of 43.2" (allowing two nod beams to be placed side by side in the slit) and a 2-pixel wide slit.  These observations were some of the earliest with MICHELLE, before proper observing protocols were established, so proved challenging to reduce.  The July 5 data were found to be superior \citep[and acquired on the same night as the images in Section \ref{images},][]{07hammel}.  However, the individual nods were poor, and the signal in the two beams differed by a factor of two (for both star and planet) due to poor nod performance and significant wind shake.   Wavelength calibration was performed by comparing the raw spectrum of Vega with atmospheric transmission from Mauna Kea, and the radiometric flux calibration was estimated using Vega.  Fourier filtering was applied to remove severe fringing from the spectra.  Unlike the 2003 and 2007 data, the slit was not aligned north-south (although it did cut across the edge of the warm south pole, which occupied just two pixels in the slit), so cannot be used to extract meridional variability of temperatures and hydrocarbons.  Instead, we scaled the averaged spectrum to be consistent with better-calibrated, disc-integrated BASS observations of the ethane emission band \citep[using the Infrared Telescope Facility,][]{06hammel} in the same time period, allowing us to treat the observations as a disc-integrated spectrum similar to the 2007 AKARI data.  

\textbf{Gemini-S (2007):} Two years after Neptune's summer solstice, we attempted to repeat these N-band observations using the 8.1-m Gemini-South observatory and the Thermal-Region Camera Spectrograph \citep[TReCS,][]{05debuizer}.  The long-slit mode (21.6") with a 0.31"-wide slit on the 8-13 $\mu$m spectrograph provides a spectral resolution of $R\approx100$ (0.1 $\mu$m near 10 $\mu$m). N-band spectra were obtained in 'Band 3' time on August 30, 2007 (05:49-06:42UT, 800 seconds of on-source time) and September 13, 2007 (01:38-02:30UT, 800 seconds on-source time) under poor weather conditions without a specific Cohen flux calibrator \citep{99cohen}.  Spectra were reduced using the Gemini IRAF MIDIR and GNIRS packages, with spectral extraction and absolute calibration performed offline with IDL.  IRAF was used to stack raw data frames and produce coadded differenced spectra of Neptune and a standard star (i.e., removing the sky background via differencing), and reference images of the sky spectra for telluric line identification and wavelength calibration.   In the absence of a Cohen standard, we used observations of HD 192947 (acquired within an hour of the Neptune spectra, with 90 seconds on-source time), a G8 class III-IV giant. We assumed a stellar temperature of 4900-K and scaled the resulting black body spectrum to match the J, H and K magnitudes as measured by 2MASS.  The resulting radiances were cross-checked against Cohen templates for standard stars of a similar spectral type to HD 192947, and then used for flux calibration.  Loss of radiance from the slit was accounted for by computing an idealised point spread function (PSF, a product of diffraction and atmospheric seeing, generated by fitting the stellar acquisition images from each run prior to the observations of Neptune) for the star at every N-band wavelength, and estimating the flux lost from the 0.31"-wide slit.  This allowed for a crude radiometric calibration of the Gemini/TReCS 2007 spectra for comparison with the earlier results. 

\subsection{Modelling AKARI}

Before considering the ground-based data, we update our previous model of AKARI spectra \citep[Fig. \ref{meanspx}(c),][]{10fletcher_akari} using Herschel-derived methane abundances in Neptune's stratosphere \citep{10lellouch}.  Furthermore, \citet{10fletcher_akari} used forward modelling with highly constrained $T(p)$ profiles, whereas we now consider full spectral inversions.  The \textit{a priori} error was tuned to fix the tropospheric temperatures, only allowing the stratospheric temperatures to vary.  Previous analyses from AKARI, ISO and Voyager favoured a quasi-isothermal stratospheric temperature structure at low pressures ($p<10 \mu$bar), so we generated a range of initial $T(p)$ profiles based on a smooth transition from the tropopause to an isotherm with temperatures in the 140-180 K range.  This range of \textit{a priori} temperatures effectively brackets the true stratospheric temperatures, and the use of multiple \textit{a priori} is designed to eliminate bias in the final solution.  Stratospheric temperatures and a scale factor for the 'Model C' ethane profile of \citet{05moses_jup} were simultaneously derived from the AKARI spectra (omitting the low-signal 9.0-11.4 $\mu$m region) for this range of \textit{a priori} isothermal structures.  Furthermore, the wavelength calibration from the AKARI pipeline had to be shifted to ensure a good fit to the data (by 0.07 $\mu$m for the methane emission at 7.7 $\mu$m and 0.04 $\mu$m for the ethane emission band at 12.2 $\mu$m).  The measurements were modelled in a disc-integrated fashion using an exponential-integral technique \citep{89goody}.  The vertical sensitivity for the AKARI SG2 module is shown in Fig. 2 of \citet{10fletcher_akari}, peaking between 0.1-1.0 mbar but with some sensitivity to $\mu$bar pressures due to a multi-lobed methane contribution function.

\begin{figure*}[tbp]
\centering
\includegraphics[width=10cm]{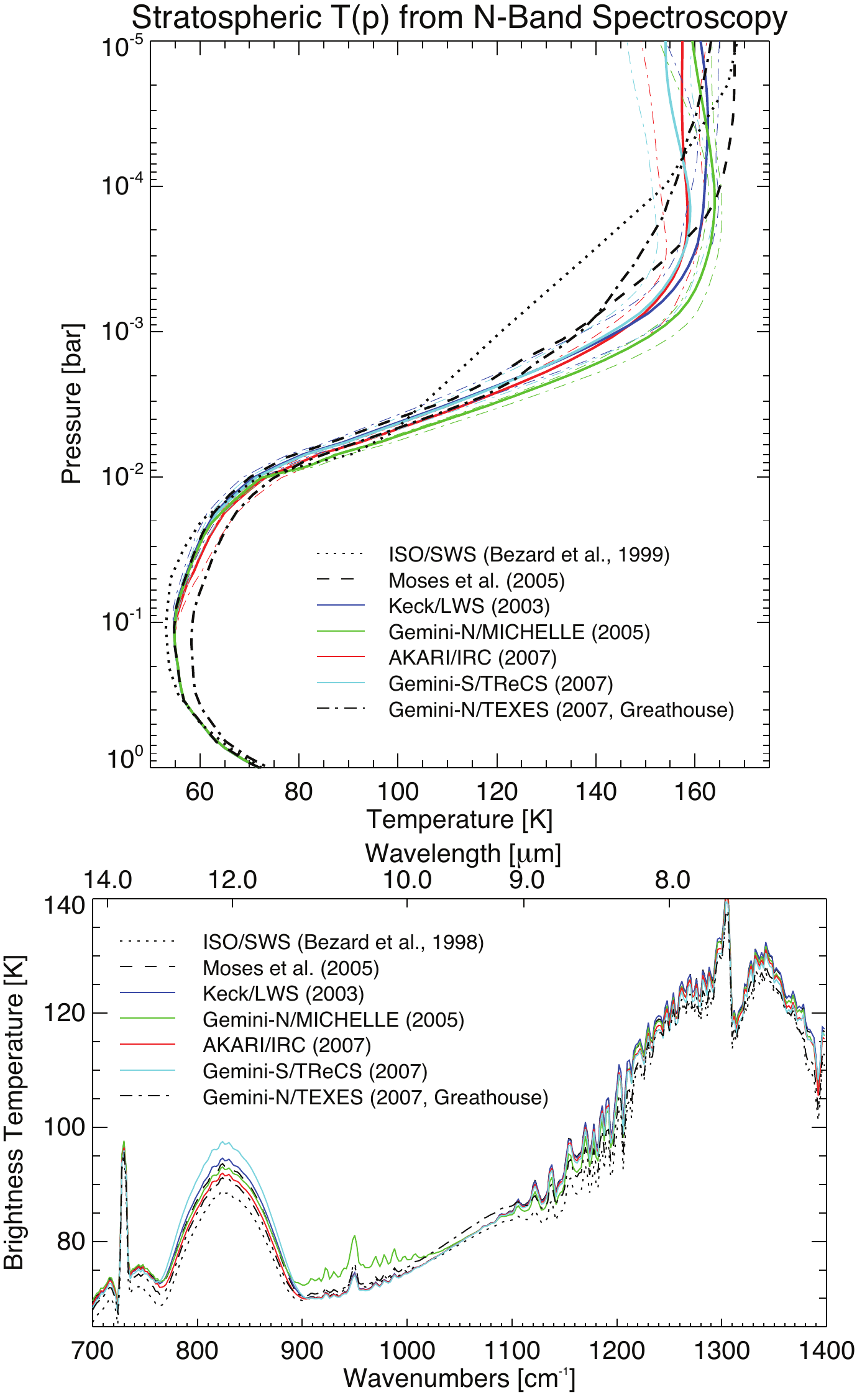}
\caption{Top panel:  Mean stratospheric temperature profiles derived from the N-band spectra considered in this study, compared to results from ISO in 1997 \citep{99bezard} and Gemini-N/TEXES in 2007 \citep{11greathouse}, plus the $T(p)$ structure assumed for the photochemical modelling of \citep{05moses_jup}.  The four profiles (combining the two different Gemini-S/TReCS observations from 2007) are given by the solid coloured lines, with estimates of uncertainty given by the colour dot-dashed lines.  The differences between our measured profiles is no larger than 6 K at 0.1 mbar and 10 K at 1 mbar (see Table \ref{tab:Nbandres}), but the overall shape remains the same.  Tropospheric temperatures are fixed to the initial \citet{05moses_jup} profile.  Bottom panel:  Forward modelled brightness temperatures at the Voyager/IRIS spectral resolution of 4.3 cm$^{-1}$ using the same $T(p)$ profiles and colour scheme as the upper panel.  For ease of comparison, all hydrocarbon abundances are set to those of \citet{05moses_jup}, and the CH$_4$ profile is based on the Herschel-derived stratospheric abundance of \citet{10lellouch}. This shows that CH$_4$ emission (unlike C$_2$H$_6$) is largely insensitive to the discrepancies between $T(p)$ profiles in the lower stratosphere.}
\label{Tprofiles}
\end{figure*}

With the Herschel-derived methane abundances, the best-fitting AKARI $T(p)$ structure is modified slightly from the results of \citet{10fletcher_akari}.  Namely, the new profile is slightly warmer at 1 mbar (143.4 K versus 139.0 K) and cooler at 0.1 mbar (158.4 K versus 165.7 K).  Fig. \ref{Tprofiles} compares the AKARI $T(p)$ to the mean temperature structures derived from the ground-based measurements, including one derived independently from Gemini-North/TEXES \citep[Texas Echelon Cross Echelle Spectrograph][]{11greathouse} observations in October 2007 (5 months after the AKARI observations).  Our derived temperature is consistent with these TEXES results at 2.1 mbar (124.3 versus 123.8 K from TEXES), and slightly warmer at 0.12 mbar (158.5 K versus 155 K from TEXES).  However, \citet{11greathouse} find a smooth $T(p)$ structure connecting these two altitudes, whereas our inversions suggest a more rapid temperature rise between 0.1-1.0 mbar.  The differences between the shapes of the AKARI and TEXES profiles, separated by only 5 months, are likely due to the added sensitivity of the TEXES inversions to the 0.1-10 mbar region.  \citet{11greathouse} required this thermal gradient between 0.1-10 mbar to reproduce limb brightening observed in TEXES observations of Neptune's H$_2$ S(1) quadrupole line.  Furthermore, fits to the H$_2$ S(0) and S(1) quadrupole lines in ISO disc-averaged spectra \citep{99bezard, 99feuchtgruber} indicated the need for a similar (albeit shallower) slope in the 0.1-10 mbar region in Fig. \ref{Tprofiles}.  As no quadrupole observations were available for this solstice study, we have used the CH$_4$ emission alone to constrain the $T(p)$ structure, which is most sensitive to the quasi-isothermal upper stratosphere at $p<1$ mbar suggested by the absence of limb brightening in methane-band images (see Section \ref{images}).  Indeed, the details of the 1-10 mbar $T(p)$ gradient have little effect on the methane emission, as indicated in the bottom panel of Fig. \ref{Tprofiles}, which compares synthetic spectra using each of the different temperature profiles.  The largest discrepancies occur in the ethane emission band near 12 $\mu$m, which has sensitivity to deeper pressures where limb brightening in ethane-band images implies the presence of the stratospheric temperature gradient. 

In the absence of the added constraint from the H$_2$ emission lines, we fit the methane and ethane emission simultaneously to derive the $T(p)$ structures in Fig. \ref{Tprofiles}.  Nevertheless, the remain differences that strongly influence the ethane derivation: the best-fit AKARI $T(p)$ suggests a C$_2$H$_6$ mole fraction of $572.0\pm21.0$ ppb at 1 mbar, whereas the cooler profile of \citet{11greathouse} requires $930^{+350}_{-260}$ ppb at 1 mbar (see Table \ref{tab:Nbandres}).  Note that the uncertainties quoted by \citet{11greathouse} are more conservative as they account for \textit{systematic} offsets of the $T(p)$ profile within the uncertainty range (we adopt the same approach in Section \ref{fixT}). Compositional uncertainties quoted here and in Table \ref{tab:Nbandres} account for \textit{randomisation} of temperatures within the Gaussian $T(p)$ error envelope (whilst also accounting for vertical smoothing), but do not consider systematic $T(p)$ offsets.  Furthermore, we do not account for uncertainties on the vertical hydrocarbon profiles, given the low spectral resolution of this dataset.  An increase in ethane by a factor of 1.7 between there AKARI observations in May and the TEXES observations in October 2007 seems unlikely, and highlights the extreme sensitivity of hydrocarbon abundances to the temperature determinations.   We assessed the necessity of allowing ethane to vary as a free parameter during the AKARI temperature retrievals.  If we were to keep ethane fixed at the abundances of \citet{05moses_jup}, the models would require a substantial temperature increase ($\Delta T \approx 10$ K) at 10 $\mu$bar pressures to fit both methane and ethane emission, something not evident in the quasi-isothermal results of previous studies.  The fits are moderately improved when ethane is also allowed to vary (goodness-of-fit $\chi^2/N=0.8$ rather than 1.1).    We conclude that a scaling of the ethane abundance is necessary to achieve an optimum fit to the AKARI spectrum.

\subsection{Keck/LWS}

Stratospheric N-band spectra from Keck (Fig. \ref{meanspx}(a)) are spatially resolved, such that spectra at a range of latitudes can be modelled with the correct emission angles.  Fig. \ref{meanspx}(a) shows how radiance in these spectra varies from north to south pole, with the coolest spectra measured in the centre of the disc. We initially fitted a mean spectrum to assess the quality of the wavelength calibration, shifting the methane and ethane bands independently until a best fit to the band location and shape was obtained.  With these modifications to the wavelength calibration, we then considered each latitude separately.  The same \textit{a priori} profile was used, fixing the tropospheric temperatures but selecting a range of isotherms (140-180 K in 5-K steps) as a starting point.  For each isothermal \textit{a priori} we retrieved temperature and an ethane scale factor for each of the 26 spectra (one for each pixel in the along-slit direction covering Neptune), and formed a stratospheric mean by averaging all resulting models together with an acceptable spectral fit between 10$^\circ$N and 50$^\circ$S.  The uncertainty on the profile shown in Fig. \ref{Tprofiles} is the standard deviation of the mean from the selected latitudes and the bracketed retrievals.  The mean $T(p)$ is moderately warmer than the AKARI profile (see Table \ref{tab:Nbandres}), but remains within 5 K of the AKARI disc-averaged results and can be considered as consistent within the errors.  

\begin{figure*}[tbp]
\centering
\includegraphics[width=15cm]{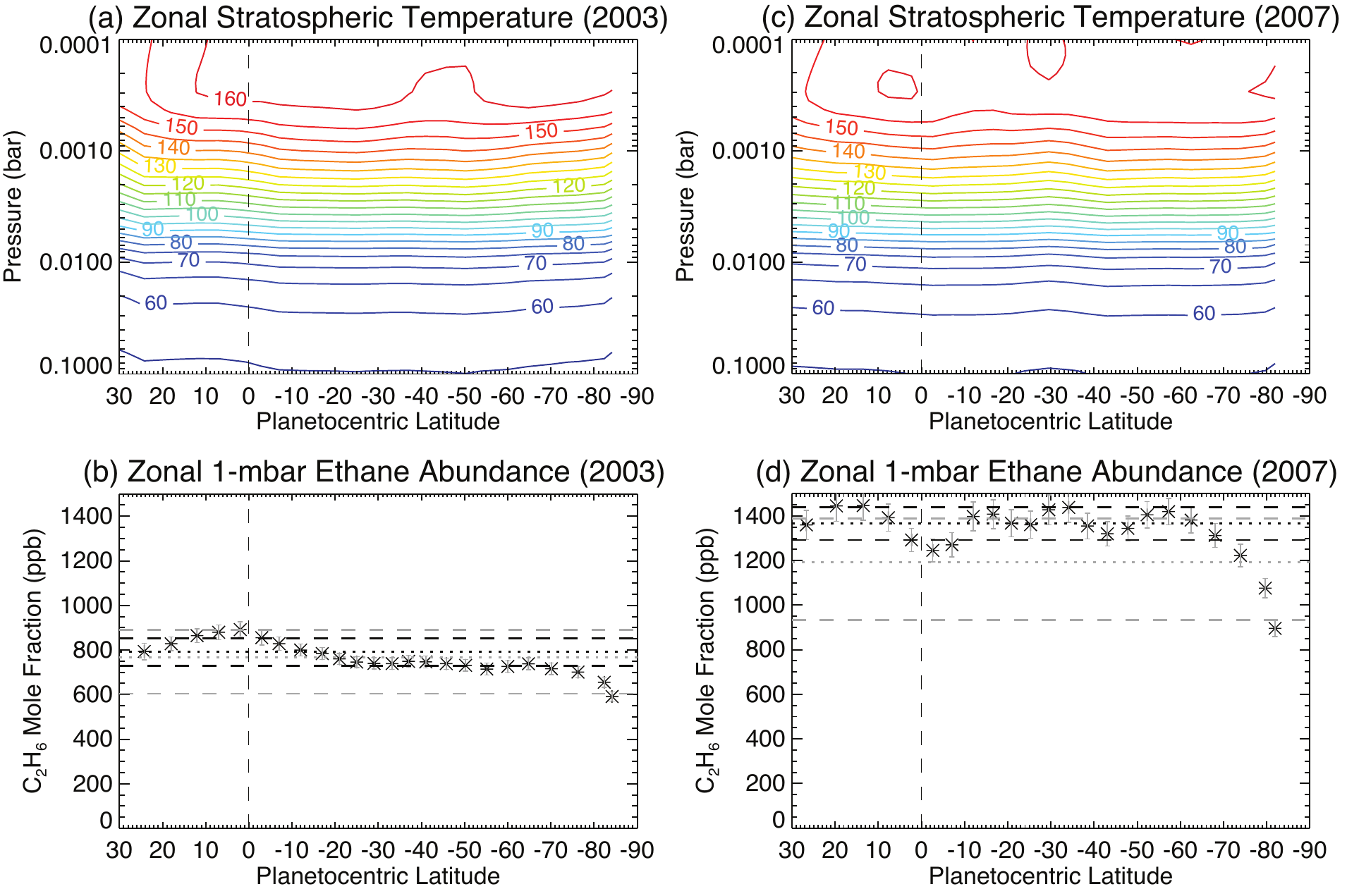}
\caption{Zonal mean temperatures and ethane abundance at 1-mbar derived from the Keck/LWS N-band spectra in 2003 and the Gemini-S/TReCS spectra in August 2007.  The cooling observed at northern latitudes is likely spurious, the effect of beam dilution at high emission angles.  Small-scale structure is limited, and the overall impression is of a meridionally-uniform thermal structure similar to that observed by \citet{11greathouse} in 2007.  The bottom panels show the abundance of ethane at the 1-mbar level.  The factor of two difference in the abundances between 2003 and 2007 is unlikely to be real, and highlights the extreme degeneracy between temperature and abundance for these low resolution spectra - the warmer stratospheric temperatures of 2003 lead to a low ethane abundance, whereas the cooler temperatures in 2007 lead to a higher ethane abundance.  The low-latitude mean abundances (Table \ref{tab:Nbandres}) are shown as horizontal dotted lines, black for the case where we simultaneously retrieve temperature, grey for the case when the temperature is fixed to the solstice mean.  Associated uncertainties on the mean abundances are given as the dashed horizontal lines using the same colour scheme, and indicate that the mean abundances between 2003 and 2007 are inconsistent with one another.  Our best fit to the 2003 Keck data suggests that the ethane abundance decreases from equator to pole, but a latitudinally-uniform distribution is still consistent with the available data. The vertical dashed line shows the equator.}
\label{meridT_c2h6}
\end{figure*}

Investigating the 2003 latitudinal retrievals in more depth, we find a surprisingly flat cross-section of stratospheric temperature (varying between 155-158 K from $20^\circ$N to $80^\circ$S at 0.5 mbar, Fig. \ref{meridT_c2h6}(a)).  \citet{11greathouse} reached a similar conclusion from their 2007 TEXES data.  Poleward of these latitudes, dilution of the planetary radiance by deep space, combined with the magnified uncertainties of retrievals at high emission angles, limits our confidence in the results.  There is a hint that northern hemisphere temperatures are cooler (147 K at $40^\circ$N compared to 156 K at $40^\circ$S at 0.5 mbar), consistent with the southern summer solstice, but Neptune's geometry is not favourable for a study of the winter hemisphere.  The cool region between 40-50$^\circ$S is reminiscent of the latitudinal profile of hydrocarbon emission measured by Voyager in 1989 \citep{91bezard}, but with a significantly smaller magnitude.  In the best-fitting case, the ethane meridional variability is more pronounced than that of temperature, with a decreasing abundance from equator to $70^\circ$S by a factor of $1.2\pm0.1$ (Fig. \ref{meridT_c2h6}(b)).  The mean profile corresponds to a best-fitting ethane abundance of $791\pm61$ ppb at 1 mbar in September 2003, larger than the AKARI ethane abundance in 2007 by a factor of 1.4.  However, when we re-ran the retrievals fixing the ethane abundance to a global mean, we were able to reproduce the Keck spectra (with a slightly poorer goodness-of-fit), suggesting that a uniform latitudinal distribution is also consistent with the data \citep[a similar conclusion was reached by ][for their 2007 TEXES data]{11greathouse}.

\subsection{Gemini-N/MICHELLE}

As the July 5, 2005 MICHELLE spectrum (Fig. \ref{meanspx}(b)) should be considered as disc-integrated, we are unable to determine the meridional variability of temperature and ethane from this dataset.  Instead, we used the exponential-integral technique to determine a globally averaged $T(p)$ profile and ethane abundance for comparison with the mean $T(p)$ from the Keck/LWS observations.  As before, a range of isothermal \textit{a priori} between 140-180 K were used to remove sensitivity to the initial profile, and a mean of all the retrievals was used as the best-fitting result.  Comparing the MICHELLE $T(p)$ to the other derivations in this work (Fig. \ref{Tprofiles} and Table \ref{tab:Nbandres}), we find that the profile was noticeably warmer in the lower stratosphere (1-10 mbar) by 7-8 K, whereas it was consistent in the stratosphere at $p<1$ mbar where the profile tends towards an isotherm.  This warmer lower stratosphere could be an artefact of the data processing techniques used - it is driven by a warmer continuum in the N-band spectrum in the lowest-signal regions (i.e., the weakest hydrocarbon lines probe the deeper stratosphere, but have a lower signal-to-noise ratio).  However, as the retrieval of temperature and ethane are intricately linked, the warmer temperatures mean that we do not require as much ethane to reproduce the spectrum, deriving a globally averaged abundance of $495\pm32$ ppb at 1 mbar, 15\% smaller than the AKARI derivation in 2007 and 37\% smaller than the Keck/LWS derivation in 2003 (Table \ref{tab:Nbandres}).  We shall return to the problem of ethane below.  

\subsection{Gemini-S/TRECS}

The 2007 Gemini-South/TReCS observations were intended to provide measurements of Neptune's stratospheric temperatures and composition shortly after solstice.  However, poor weather degraded seeing and the absence of a Cohen standard worsened calibration accuracy compared to the September 2003 Keck/LWS results.  N-band spectra were analysed in the same fashion, starting with a mean spectrum to assess wavelength calibration modifications, then fixing the tropospheric temperatures to our \textit{a priori} and considering a range of isothermal stratospheric temperature structures to bracket the result.  Twenty-six spectra were analysed to cover the planet from $55^\circ$N to $80^\circ$S for both the August and September observations, shown in Fig. \ref{meanspx}(d-e).  We find that the mean $T(p)$ profile is consistent with that from AKARI in the same year and slightly cooler than that from Keck four years earlier, but all are consistent to within the retrieval uncertainties (Fig. \ref{Tprofiles}).  Slight differences are obtained from the August 2007 and September 2007 results (see Table \ref{tab:Nbandres}), but these are within the absolute errors of 3-5 K estimated for this dataset.  This validates our approach of calibrating by scaling a stellar blackbody to match 2MASS J, H and K band fluxes.  Just as for the Keck results in 2003, we observe only small contrasts in the north-south temperatures in the stratosphere (Fig. \ref{meridT_c2h6}(c)), and no evidence for a cool region between 40-50$^\circ$S.  The cool northern stratosphere in Fig. \ref{meridT_c2h6}(c) is likely an artefact of dilution of the planetary radiance close to the limb.  

However, the meridional ethane distribution in Fig. \ref{meridT_c2h6}(d) no longer shows any equator-to-pole contrasts, and instead reveals variability that was not seen in the Keck 2003 data.  For example, the raw TReCS spectra show a dark band at the equator at 12.3 $\mu$m that is being interpreted by our retrieval as a decrease in equatorial ethane by 15\% compared to $\pm20^\circ$ latitude.  A dark band was not observed in the raw Keck/LWS data in 2003, nor is there any evidence for this sort of structure in the acquisition images of Neptune obtained in the minutes before the spectroscopy.  The same structure was seen in data on both August 30th and September 13th 2007, but Neptune was in exactly the same location on the detector array.  A further dark band is observed in the spectral dimension for all latitudes (the 'notch' observed in Fig. \ref{meanspx}(d-e) in the centre of the ethane band), which cannot be explained in terms of terrestrial or neptunian spectra.  It seems likely that the Band-3 (i.e., low-priority) nature of the TReCS data has introduced artefacts that could have been removed if a second dataset had been obtained, shifted slightly on the array.  

More importantly, despite the similarities of the temperature profile between the 2003 and 2007 results, the estimation of the ethane abundance varies substantially (Table \ref{tab:Nbandres}).  The mean C$_2$H$_6$ across latitudes $10^\circ$N to $50^\circ$S in August 2007 was $1366\pm74$ ppb, an increase by a factor of $2.0\pm0.1$ with respect to 'Model C' of \citet{05moses_jup}.  A similar enhancement was found in the September 2007 data, with discrepancies related to the slightly different temperature profiles retrieved from these two datasets (Table \ref{tab:Nbandres}).  The high ethane abundances are being driven by the contrast between the bright ethane band and the dim methane band observed in in both the August and September TRECS spectra in Fig. \ref{meanspx}(d-e), and we investigate ethane variability further in the next subsection.

\begin{sidewaystable}[htp]  
\caption{Summary of N-Band Spectroscopy Results: These models all use the same CH$_4$ abundances as \citet{10lellouch}, but the Gemini-N/TEXES temperatures and ethane abundances were kindly provided by \citet{11greathouse}.   Notes:  (1) Absolute errors on temperatures are approximately 3-5 K.  (2) The scale factor is given relative to 'Model C' of \citet{05moses_jup}, which has 676 ppb at 1 mbar.  (3) With the exception of the result from \citet{11greathouse}, uncertainties on the ethane abundances and scale factors in these columns are solely due to fitting uncertainties and randomised temperature errors, and do account for a systematic shift of $T(p)$ within the error envelope nor uncertainties on the vertical ethane profile.  (4) The final two columns present results for a fixed $T(p)$ profile and provide more conservative ethane uncertainties that systematically shift the $T(p)$ within the $1\sigma$ error range on the fixed-$T(p)$ profile.  These uncertainties are comparable in size to those quoted by \citet{11greathouse}.  }
\begin{tabular}{cccccccc}
\hline
\hline
Data Source & Date & T (1.1 mbar)$^1$ & T (0.1 mbar)$^1$ & Ethane at 1 mbar$^3$ & Scale Factor$^{2,3}$ & Fixed T C$_2$H$_6$$^4$ & Fixed T Scale$^4$\\ 
\hline\\
Keck/LWS & 2003-09 & 142.5 & 161.9 & $791\pm61$ ppb & $117\pm9$\% & $767_{-163}^{+122}$ ppb & $113_{-24}^{+18}$\% \\
Gemini-N/MICHELLE & 2005-07 & 150.7 & 164.0 & $495\pm32$ ppb & $70\pm4$ \% & $609_{-136}^{+104}$ ppb & $90_{-20}^{+15}$\% \\
AKARI/IRC & 2007-05 & 143.4 & 158.4 & $572\pm21$ ppb & $85\pm3$\% & $528_{-116}^{+89}$ ppb &  $78_{-17}^{+13}$\%\\
Gemini-S/TReCS & 2007-08 & 141.4 & 158.8 & $1366\pm74$ ppb & $202\pm11$\% & $1192_{-258}^{+196}$ ppb & $176_{-38}^{+29}$\% \\
Gemini-S/TReCS & 2007-09 & 140.3 & 162.0 & $1223\pm65$ ppb & $181\pm10$\%& $1220_{-264}^{+201}$ ppb & $180_{-39}^{+30}$\% \\
Gemini-N/TEXES & 2007-10 & 132.8 & 154.7 & $930^{+350}_{-260}$ ppb & $137^{+52}_{-38}$\% & - & - \\
\hline\\
\end{tabular}

\label{tab:Nbandres}
\end{sidewaystable}

\subsection{Variability of Stratospheric Ethane}
\label{fixT}
\begin{figure}[tbp]
\centering
\includegraphics[height=17cm]{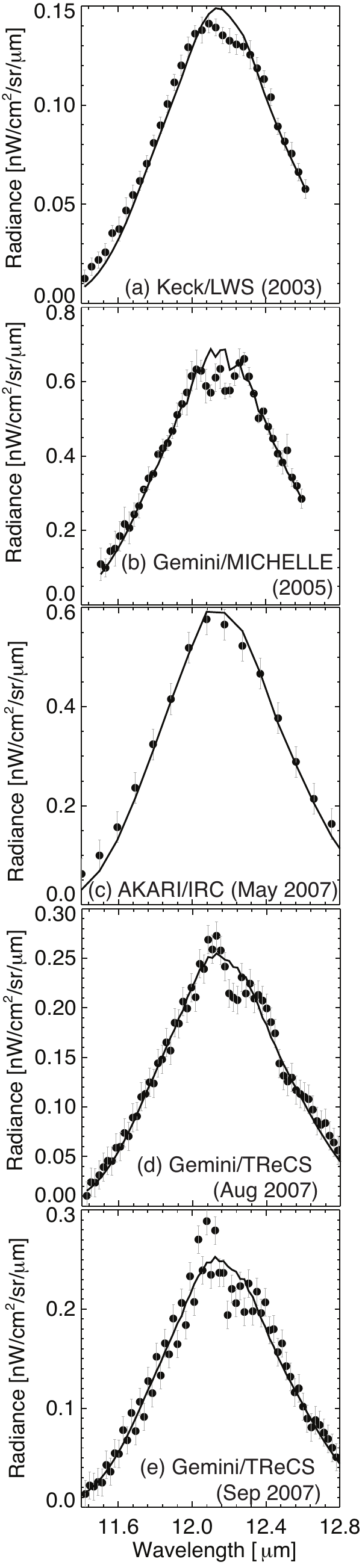}
\caption{Example fits to the ethane emission band when we fix stratospheric temperatures to a mean $T(p)$ over the 2003-2007 period.  Panels (b) and (c) are disc-averaged spectra, the remaining panels are averages along the slit for emission angles smaller than $50^\circ$.  The broad shape of the band is reproduced within the measurement uncertainties, but the source of structure ('notches') within the centre of the C$_2$H$_6$ band in Gemini data is uncertain, although both use the same detector array.  A similar issue occurs near the ethane peak in the 2003 Keck observation (a).}
\label{c2h6fit}
\end{figure}

The derived abundances of ethane in Table \ref{tab:Nbandres} are intimately tied to the details of their respective temperature structures.  Even small differences between $T(p)$ profiles can lead to substantial changes in the amount of ethane required to reproduce the 12.3-$\mu$m emission features.  Given the similarity of the temperature profiles in Fig. \ref{Tprofiles} from 2003 to 2007,  we now fix the $T(p)$ profile to a single mean for all dates so that we can explore the yearly consistency of the ethane peak without the complicating influence of the uncertain $T(p)$ profile.  Fixing the $T(p)$, we simply scale the C$_2$H$_6$ profile of \citet{05moses_jup} to fit the 11.4-12.6 $\mu$m ranges of the Keck, Gemini-N, Gemini-S and AKARI datasets, showing representative model fits in Fig. \ref{c2h6fit} and the corresponding ethane abundances in Table \ref{tab:Nbandres}.  Here we adopt the more conservative approach of \citet{11greathouse} and systematically shift the $T(p)$ profile within the $1\sigma$ error envelope to determine the uncertainties on the ethane scale factor.  The Keck/LWS, Gemini-N/MICHELLE and AKARI/IRC ethane results are now consistent with each other within these uncertainties, and tentatively suggest a decreasing globally-averaged ethane abundance from 2003 to mid-2007.  The Gemini-S/TReCS results in August and September 2007 are considerably larger, and more consistent with the Gemini-N/TEXES findings in October 2007 \citep{11greathouse}. 

The increase in ethane emission in late 2007 is hard to understand given the invariance of stratospheric temperature if the calibration and interpretation of the N-band spectra is robust.  However, several factors may complicate the interpretation of ethane variability on Neptune:
\begin{enumerate}
\item Small amounts of water ice in cirrus clouds would attenuate 12-$\mu$m transmission of either Neptune or the standard star, whereas 8-$\mu$m transmission would be largely unaffected.  Calibrated radiances at 12 $\mu$m could be corrupted if the weather conditions were less than perfect, whereas the methane emission and derived $T(p)$ profiles would be unaffected, leading to a seemingly-random variation of the ethane emission.
\item The absolute calibration technique for the Gemini-S data that, in the absence of a suitable divisor, were scaled to match the observed bright star to J, H and K band fluxes measured by 2MASS, may be in error. 
\item Notches observed in the centre of the ethane band (Fig. \ref{c2h6fit}), possibly related to the Si:As Raytheon detector used by both the TReCS and MICHELLE instruments, could be symptomatic of a larger problem with the shape of the C$_2$H$_6$ band being measured. 
\item Contribution functions near the ethane band in Fig. \ref{refspx} show (a) a multi-lobed structure with some contribution from tropospheric temperature and (b) that methane emission probes slightly higher altitudes than ethane emission.  If temperatures are significantly altered in the troposphere or lower stratosphere, for example by upward propagation of waves, then the CH$_4$ emission might not provide sufficient information for a unique $T(p)$ determination.  This is highlighted by the modified lower stratospheric temperatures of $T(p)$ profiles that take H$_2$ quadrupole emission into account \citep{99bezard,11greathouse}. 

\end{enumerate}

It seems likely that the variable ethane emission we observe is either due to poorly-quantified telluric weather or to changes in Neptune's lower stratospheric temperature beyond our capabilities to measure using low-resolution spectroscopy.  Large variations in ethane emission, apparently divorced from variability in methane emission, have been previously reported \citep{06hammel}, and the amplitude of the variations reported here is consistent with their study.  A continuous campaign of spectral observations over multiple years, preferably outside of the complicating influence of the terrestrial atmosphere, would be required to understand the long term variability of methane and ethane emission on Neptune.

In summary, we find variations in the 0.1-10.0 mbar temperatures retrieved from the solstice-era N-band spectroscopy (Fig. \ref{Tprofiles} and Table \ref{tab:Nbandres}), but these variations are smaller than the uncertainties related to radiometric calibration and those associated with retrieval degeneracy.  Nevertheless, retrievals from independent observatories all support (i) a latitudinally uniform distribution of stratospheric temperatures despite southern solstice conditions;  and (ii) a quasi-isothermal structure at altitudes above the 0.1-mbar pressure level.  The close correspondence of the $T(p)$ profiles derived from different instruments under vastly different observing conditions is encouraging, and suggest that stratospheric temporal variability, if present, is $<\pm5$ K at 1 mbar and $<\pm3$ K at 0.1 mbar.

\section{Tropospheric Q-Band Spectroscopy}
\label{tropos}

Terrestrial water vapour obscures large regions of the 17-25 $\mu$m Q band, limiting our ability to calibrate ground-based Q-band observations unless the observing nights are particularly dry.  Keck/LWS observations provided a disc-averaged Q-band spectrum that can be directly compared to the Voyager/IRIS results from 1989.   In this section, we use the stratospheric temperatures from Section \ref{stratos} as a prior for fitting both Voyager/IRIS zonal mean spectroscopy in 1989 and Keck/LWS Q-band spectra from 2003.  The resulting zonal mean thermal structure will be used in Section \ref{images} to generate synthetic images for comparison with observations.

\subsection{Reanalysis of Voyager/IRIS Spectra}

\begin{figure}[tbp]
\centering
\includegraphics[width=6cm]{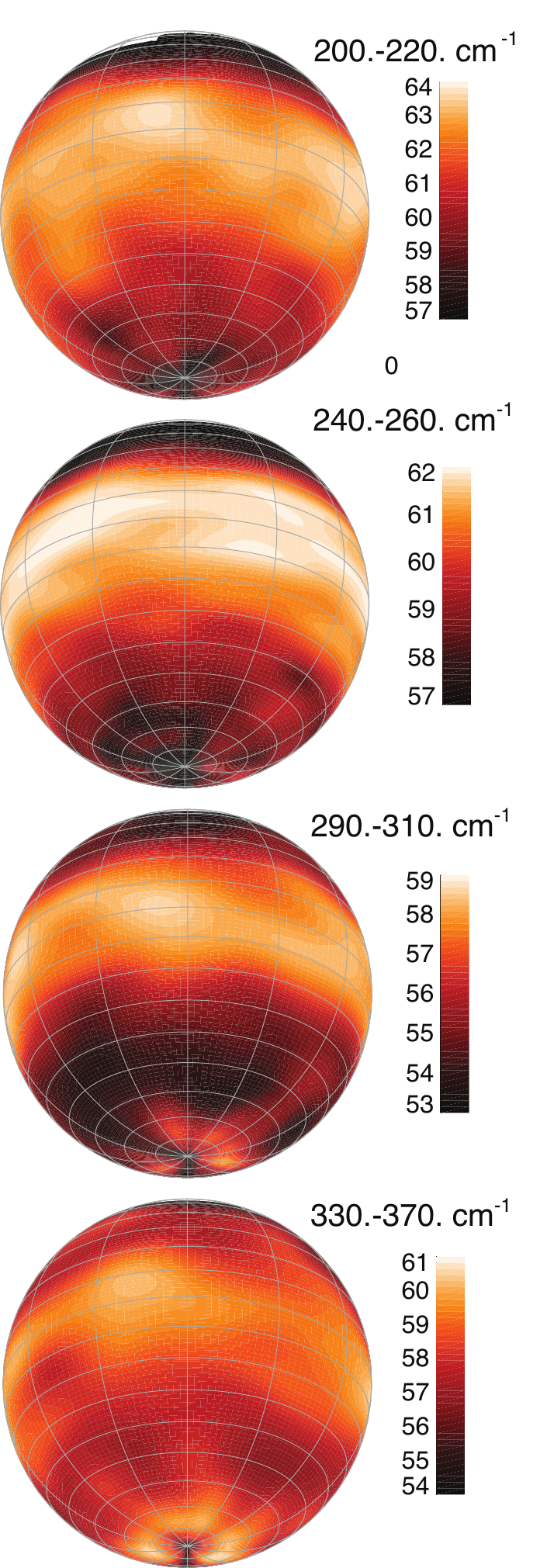}
\caption{Neptune's appearance in 1989, from spherical reprojection of Voyager-IRIS brightness temperatures in four wavelength ranges.  The scale bar shows brightness temperatures in kelvin.  The irregular distribution of spectra across Neptune's disc necessitated interpolation and smoothing to remove artefacts.  Projection effects and a lack of data lead to considerable uncertainties at high latitudes (particularly the dark spot right at the south pole, as no data were acquired at these high southern latitudes). The equator to pole decrease in brightness at 200-260 cm$^{-1}$ is consistent with limb darkening, whereas the slight increase in polar brightness at 290-350 cm$^{-1}$ is due to limb brightening (sensitivity to the lower stratosphere near the peak of the broad S(0) absorption at 354 cm$^{-1}$).  The warm equator is present for all Q-band wavelengths between 25-50 $\mu$m (where IRIS SNR$>1$). }
\label{iris_spheres}
\end{figure}

\begin{figure}[tbp]
\centering
\includegraphics[width=6cm]{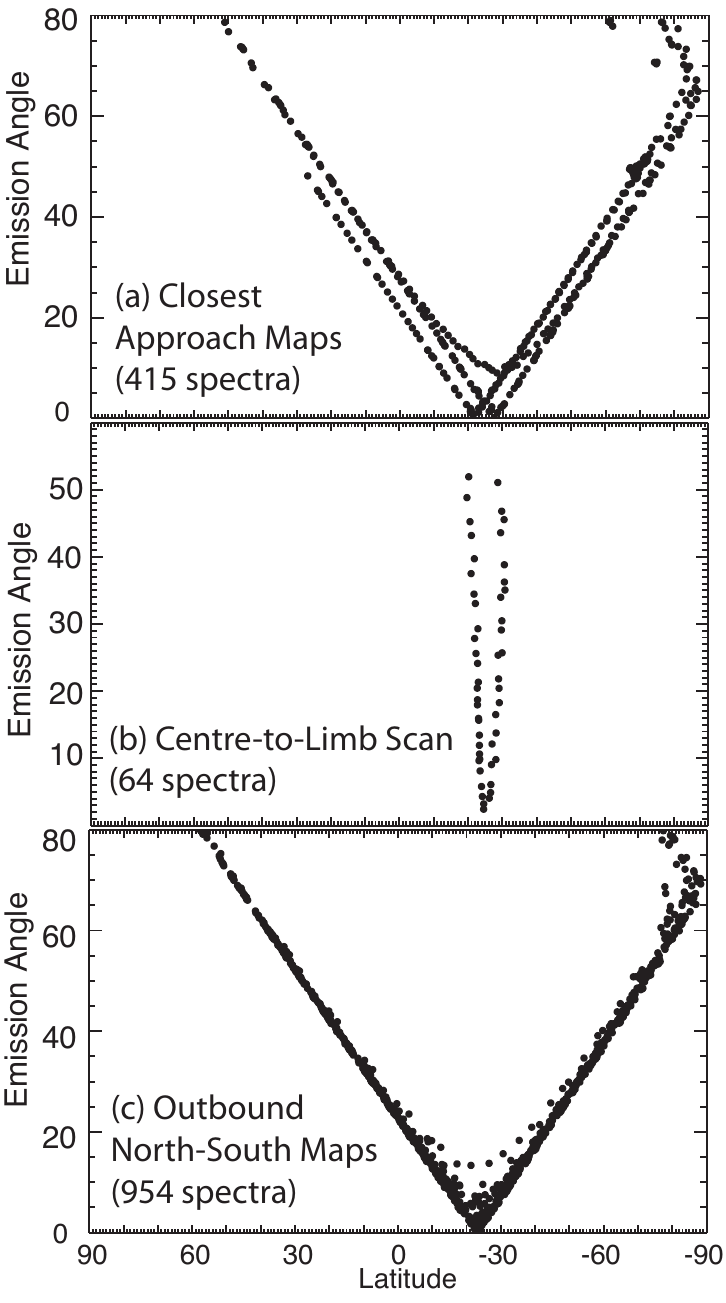}
\caption{Emission angles for Voyager-2/IRIS data extracted from the Planetary Data System.  Spectra were broadly classed as 'closest approach' and 'outbound', although the former maps consisted of a number of different data types (meridional scans, latitudinal stares, etc.).  Outbound maps have been previously studied by \citet{89conrath} and \citet{98conrath} as they contained the most contiguous spectra.  The centre-to-limb scan centred at 20-30$^\circ$S has not been published previously.  }
\label{iris_data}
\end{figure}

\begin{figure}[tbp]
\centering
\includegraphics[width=6cm]{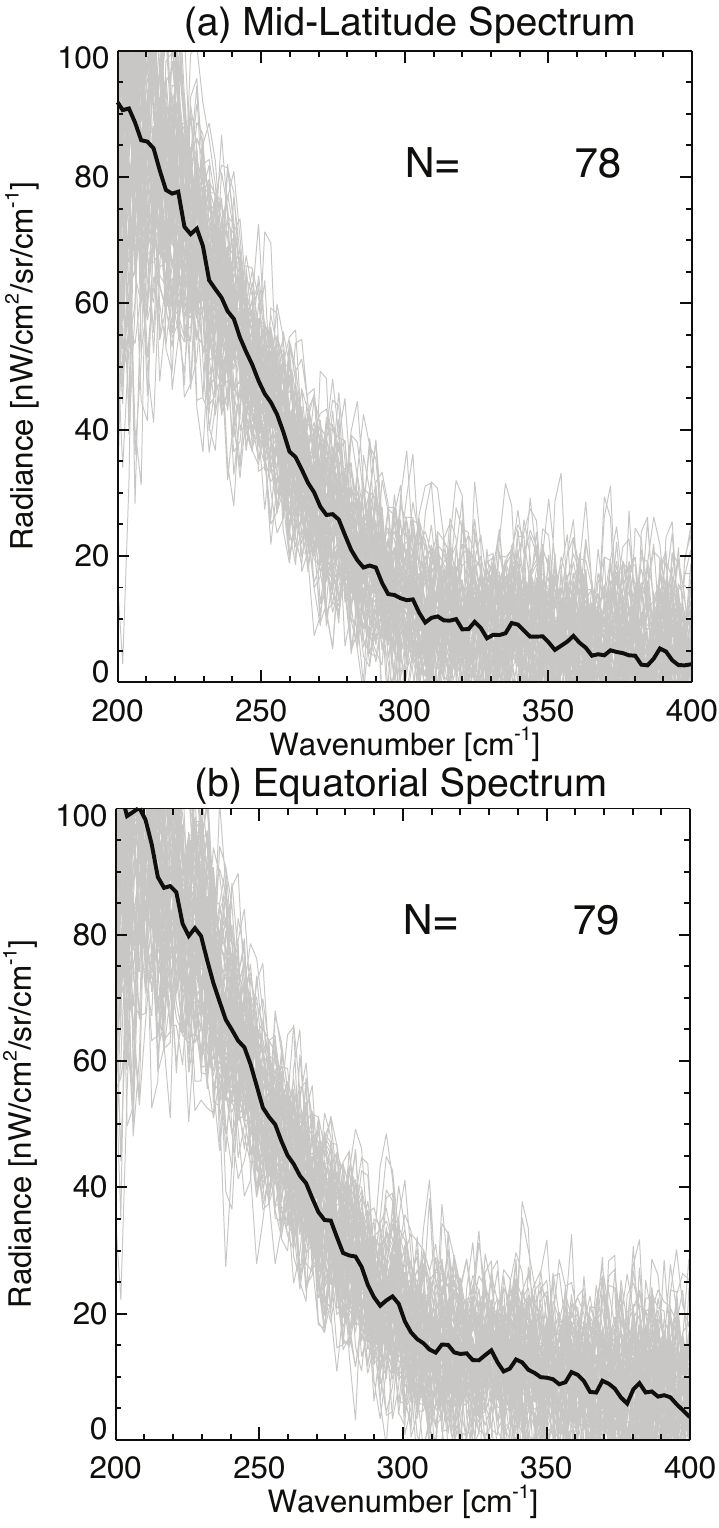}
\caption{Two examples of IRIS spectra at southern mid latitudes and the equator.  The solid black curve is the mean radiance (once low-quality spectra have been filtered out), the grey envelope show the range of the data contributing to the mean.  The number of spectra in the mean, $N$, is between 60 and 80 for each latitude bin.  The equatorial spectrum is slightly warmer than the mid-latitude example.}
\label{iris_spectra}
\end{figure}

Voyager 2 launched on August 20, 1977, and encountered Neptune twelve years later, on August 25, 1989 with a closest approach of 1.3 R$_N$.  Voyager/IRIS \citep[Infrared Spectrometer and Radiometer,][]{80hanel} data was extracted from the VG2001 archive on the Planetary Data System\footnote{http://pds-rings.seti.org/vol/VG\_2001/NEPTUNE/}.  The dataset extends over the three days surrounding closest approach (August 23-26 1989, spacecraft flight data system (FDS) count 1133200-1144200).  A global average of all these data provided 1916 spectra in total, centred at 32$^\circ$S planetographic latitude with a mean emission angle of 35$^\circ$.  The spectra were further divided (Fig. \ref{iris_data}) into maps between counts of 136227-1140951 (415 spectra in total, shortly after closest approach from distances of $0.85-1.3\times10^6$ km) and maps between counts of 1141600-1144200 (954 spectra from distances between $1.3-2.5\times10^6$ km), with the latter having a more uniform distribution of emission angles and target distances.  The appearance of Neptune at four wavelengths sampled by IRIS is shown in Fig. \ref{iris_spheres} to demonstrate the spatial variation of the 25-50 $\mu$m emission.  The north-south scanning spectra were binned on a 5$^\circ$ latitude grid with bins 10$^\circ$ wide (including 60-80 spectra in each average, as shown in Fig. \ref{iris_spectra}).  Within each bin, only those spectra with emission angles within 10$^\circ$ of the mean were allowed to contribute to the mean.  Furthermore, only spectra with a `zero' reject code  were included (i.e., the interferogram was considered to be good).  Although data are provided between 150-900 cm$^{-1}$ in 2.15 cm$^{-1}$ intervals (i.e., Nyquist sampling of the 4.3 cm$^{-1}$ IRIS resolution), only the 200-400 cm$^{-1}$ range produced usable spectra with a signal-to-noise ratio exceeding unity.   Spectra were filtered to remove any large ripples (from spikes or waves in the interferograms), and are specified by their FDS count (each count represents one 48-second interferogram).   An additional centre-to-limb scan of Neptune's Q-band at 20-30$^\circ$S (64 spectra) will also be presented.  

\subsubsection{Tropospheric Temperature and Para-H$_2$}
\begin{figure*}[tbp]
\centering
\includegraphics[width=13cm]{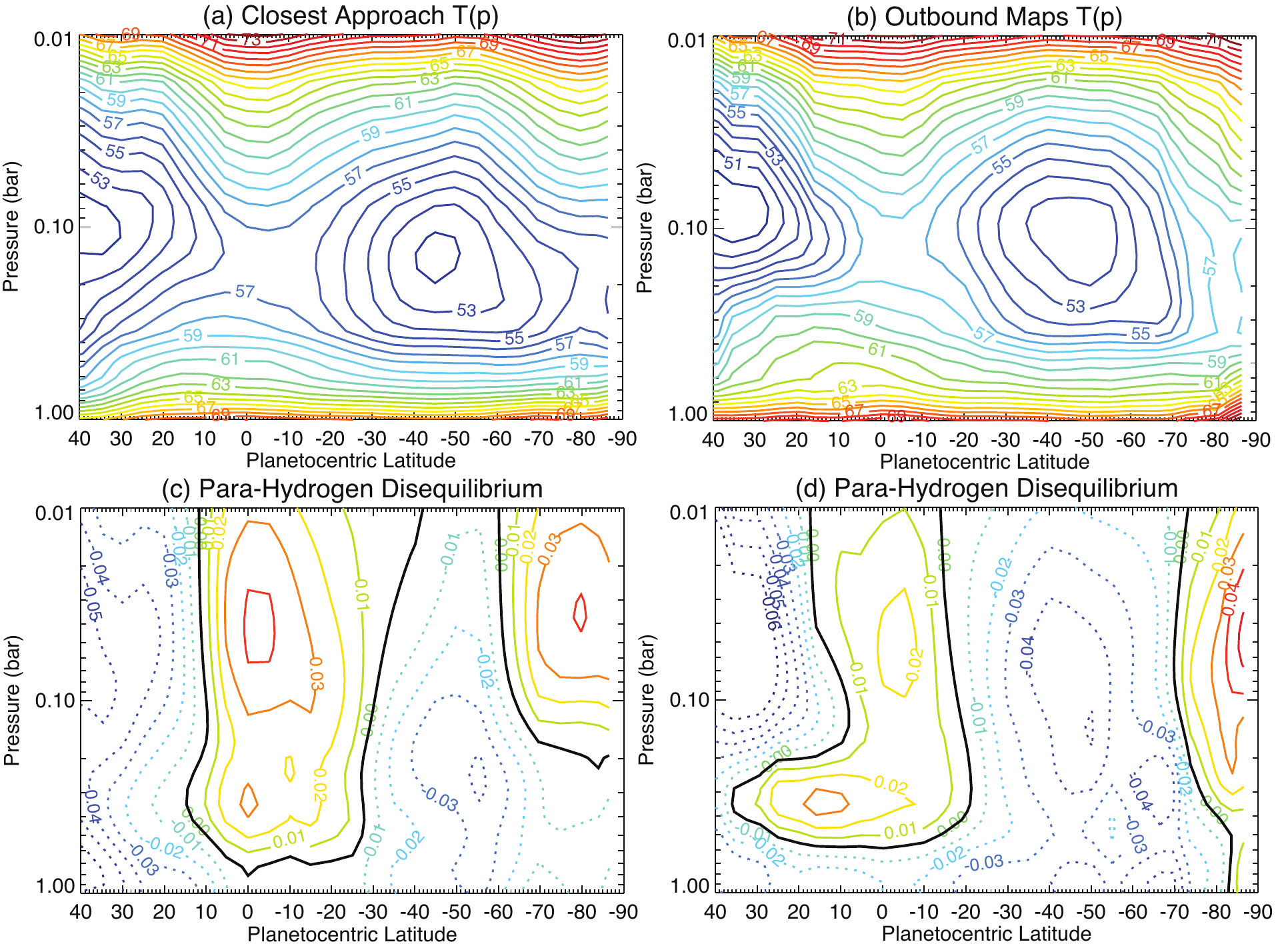}
\caption{Comparison of Neptune's temperatures and para-H$_2$ disequilibrium between the closest-approach and outbound IRIS maps, demonstrating the similarities between the two.  Temperature contours are given at 1-K intervals, and demonstrate the cool mid-latitudes and warm equator and pole.  Although these figures cover the 10-1000 mbar range, reliable information on temperature and para-H$_2$ come only from the 70-800 mbar and 200-800 mbar ranges, respectively. Outside of these domains the profiles represent a smooth relaxation to our \textit{a priori}.  Super-equilibrium conditions ($f_p>f_{eqm}$, solid contours in panels (c) and (d)) prevail at the equator and high southern latitudes, indicative of atmospheric subsidence; whereas sub-equilibrium conditions are seen at mid-latitudes in both hemispheres ($f_p<f_{eqm}$, dotted contours), coincident with the coldest atmospheric temperatures. Differences in the para-H$_2$ profiles are unlikely to be realistic, and indicate the limitations  and uncertainty of non-unique retrievals from noisy data.}
\label{iris_Tp_fp}
\end{figure*}

Starting from an \textit{a priori} profile consisting of the mean stratospheric $T(p)$ and the tropospheric profile of \citet{05moses_jup}, we retrieved the latitudinal variation of tropospheric temperatures and the fraction of ortho-to-para hydrogen from the 200-400 cm$^{-1}$ collision-induced continuum on a 5$^\circ$ latitude grid for both Voyager mapping sequences.  The sensitivity of our retrieval to the \textit{a priori} temperature and para-H$_2$ fractions was tested by running multiple retrievals with different smoothing constraints.  Meridional temperatures derived from the two mapping sequences compare favourably in Fig. \ref{iris_Tp_fp}, producing similar meridional trends.  Although the domain in this figure extends from 10-1000 mbar, the IRIS spectra only provide temperature information on the 70-800 mbar range.  Temperatures above and below this domain are from smooth relaxation to our \textit{a priori} profile. The $T(p)$ profiles compare well with those of Fig. 6 of \citet{98conrath} - at the tropopause we see a minimum temperature of 51 K at 45$^\circ$S; equatorial temperatures in the 56-57 K range; northern mid latitude temperatures decreasing towards 51 K; and south polar temperatures approaching 56 K.  The hint of cooling right at $85^\circ$S is likely to be a spurious projection effect.  Temperatures at 1 bar  approach 68 K, consistent with the radio occultation profiles of \citet{92lindal_nep}.  In the lower stratosphere we see equatorial temperatures of  60-63 K in the 30 mbar region, consistent with \citet{98conrath} but slightly higher than \citet{92lindal_nep}.

The degree of para-H$_2$ disequilibrium looks rather different from the calculations in Fig. 9 of \citet{98conrath}.  Indeed, the distribution of para-H$_2$ is more sensitive to the assumptions made in the retrievals, and robust information is only available in the 200-800 mbar range in Fig. \ref{iris_Tp_fp}.  Our new results, using updated CIA calculations, appear to be more symmetric about the equator than those of \citet{98conrath}, consistent with a distribution driven by weather-layer dynamics rather than seasonal effects.  \citet{98conrath} measured sub-equilibrium conditions ($f_p<f_{eqm}$, indicative of upwelling of low-$f_p$ air) from 0-50$^\circ$S, super-equilibrium conditions ($f_p>f_{eqm}$, indicative of subsidence) poleward of 50$^\circ$S and throughout the northern hemisphere.   Our analysis of the same spectra suggest super-equilibrium at the equator ($\pm20^\circ$) and at high southern latitudes, with sub-equilibrium conditions at mid-latitudes in both hemispheres.  The source of the discrepancy with \citet{98conrath} is unclear, but may relate to our use to updated collision-induced absorption spectra; differing assumptions about the degree of smoothing required; or simply the difficulty in extracting multiple variables from a continuum.  This is consistent with a meridional circulation, with cold air rising at mid-latitudes and subsiding at both the poles and the equator.  It is interesting to note that this circulation is in the opposite sense to that of Saturn, where sub-equilibrium conditions are observed at the equator and seasonal effects produce a north-south asymmetry in the para-H$_2$ fraction \citep{07fletcher_temp}.  The results from the two mapping sequences in Fig. \ref{iris_Tp_fp} do differ slightly, reflecting the sensitivity to the retrieval assumptions and the fact that the closest-approach maps are noisier because fewer spectra were used, implying that the retrieval is more weighted to the \textit{a priori} assumptions.  Nevertheless, the meridional trends in para-H$_2$ disequilibrium are similar in both maps.  The meridional temperature and para-H$_2$ variations from Voyager/IRIS will be used in Section \ref{images} to generate synthetic images of Neptune for comparison with the data. 

\begin{figure}[tbp]
\centering
\includegraphics[height=16cm]{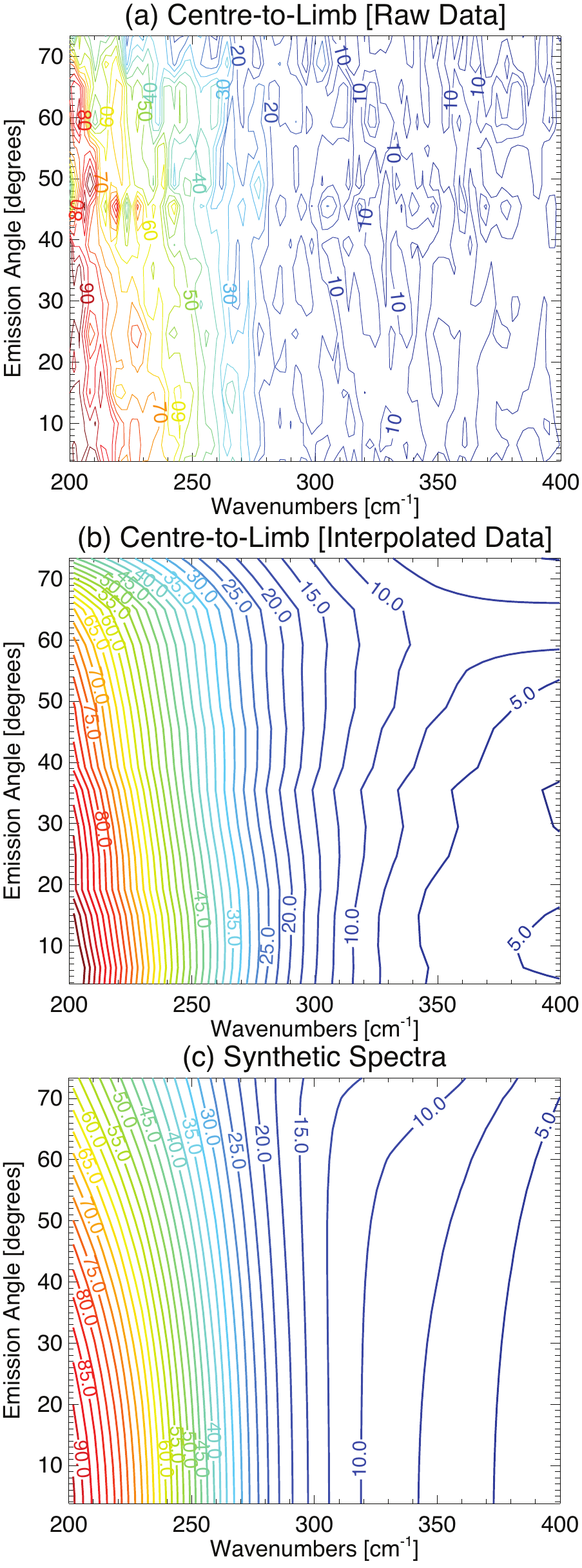}
\caption{Centre-to-limb variation of Neptune's emission measured by IRIS between 200-400 cm$^{-1}$.  The upper panel shows the noisy raw data, produced from 64 spectra by averaging onto a 5$^\circ$ emission angle grid.  The data is smoothed by spline interpolation in panel (b), which facilitates easier comparison with synthetic spectra in panel (c).  The synthetic spectra were created using the best-fitting IRIS and stratospheric temperature and para-H$_2$ profiles, sampled on the same emission angle grid.  The model is capable of reproducing the limb-darkening in the 200-300 cm$^{-1}$ region, and also the slight limb brightening visible for longer wavenumbers (due to sensitivity to the lower stratosphere near the S(0) absorption).  }
\label{irisc2l}
\end{figure}

Finally, additional constraints on the vertical tropospheric $T(p)$ structure can be deduced from the centre-to-limb variation of the H$_2$-He emission.  The Voyager/IRIS centre-to-limb scan was acquired near to closest approach, scanning from east to west along a latitude circle of 20-30$^\circ$S.  The 64 spectra were binned onto a 5$^\circ$ emission angle grid, with 10-12 spectra contributing at low emission angles and 2-4 spectra at high emission angles.  These noisy data were smoothed using the spline interpolation algorithm of \citet{07teanby_splines} to facilitate easier comparison with synthetic spectra.  Spectra were calculated using the best-fitting IRIS $T(p)$ for 25$^\circ$S and the stratospheric $T(p)$ profile at a range of emission angles, and are compared to the IRIS data in Fig. \ref{irisc2l}.  The model does an excellent job at reproducing the observed radiances and the centre-to-limb variation, including the limb darkening at the high-wavelength end and the slight limb brightening between 300-400 cm$^{-1}$ (this was also evident in the polar brightening in Fig. \ref{iris_spheres}).   The limb brightening occurs because the continuum near the S(0) H$_2$ collision-induced absorption (354 cm$^{-1}$) produces enough opacity that spectra are sensitive to the rising temperatures of the lower stratosphere (see Fig. \ref{refspx}).

\subsection{Keck/LWS Q-Band Spectra}

The Keck/LWS Q-band spectrum was acquired on September 6, 2003 (08:39-09:28 UT, total integration time of 768 seconds) between 17.6 and 22.4 $\mu$m, and full details of the spectral acquisition and reduction are provided by \citet{13depater}.  The sit was oriented north to south on Neptune's disc, and the 128 pixels of the LWS detector provided a spectral resolution of $R\approx100$ in the Q band.  Standard stars HD198542 and HD188154 \citep{99cohen} were used for telluric corrections and photometric calibration.  Unlike the N-band spectrum, we did not detect any north-south variability along the slit, even though the diffraction-limited spatial resolution (0.44-0.56 arcseconds over the Q-band, omitting the effects of variable seeing) should have been sufficient to resolve equator-to-pole contrasts on the 2.3" Neptune disc (see, for example, the images in Section \ref{images} using the same observatory).  We therefore converted the measurements into a disc-integrated spectrum by multiplying the summed spectrum by a factor of $4.05\pm0.06$ (based on superimposing the LWS slit onto Neptune images to estimate that the slit receives $24.7\pm0.4$\% of the total intensity from the planet's disc).  

\begin{figure}[tbp]
\centering
\includegraphics[width=8cm]{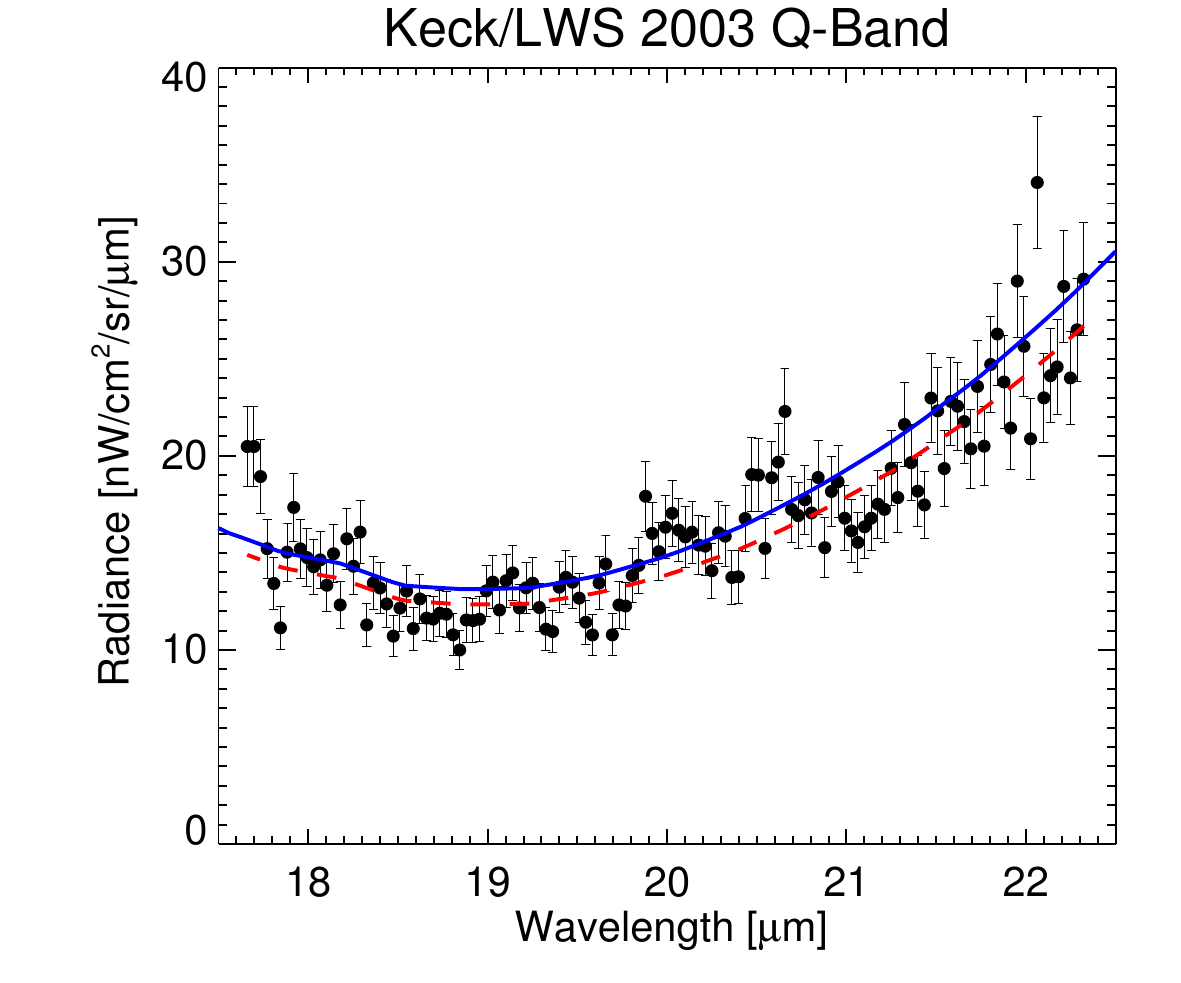}
\caption{Disc-integrated Q-band spectrum of Neptune between 17.6-22.4 $\mu$m measured by Keck/LWS on September 6th 2003 (dots with 10\% error bars).  The spectrum is expected to be featureless in this range (smooth collision-induced absorption due to H$_2$ and He), so the features are likely to be artefacts due to imperfect removal of telluric water vapour.  The solid line is a forward model based on the best-fitting Voyager/IRIS tropospheric temperature profile.  The dashed line is a full retrieval model, but the difference is insignificant.  We conclude that the globally-averaged tropospheric temperature has varied by no more than 0.7 K since the Voyager/IRIS observations in 1989.}
\label{keckQ}
\end{figure}

Having reproduced the tropospheric temperature variations observed by Voyager/IRIS \citep{89conrath, 98conrath}, we now search for consistency with the disc-integrated Keck/LWS 17-24 $\mu$m spectrum (2003) in Fig. \ref{keckQ}.  Any discrepancies between the global-mean $T(p)$ from Voyager and the Keck/LWS spectrum would be suggestive of seasonally evolving temperatures over the 14 years between 1989 and 2003, even though the Keck results sample a shorter wavelength region than the IRIS results (17-24 $\mu$m rather than 25-50 $\mu$m).  As this region of the spectrum is expected to be featureless, we believe that the small variations observed in Fig. \ref{keckQ} are due to imperfect removal of telluric absorption features during the Q-band calibration process.  

The mean $T(p)$ and para-H$_2$ fraction from Voyager were used to simulate a disc-averaged spectrum for comparison with the Keck/LWS Q-band data.  Given the sub-spacecraft latitude of 23$^\circ$S during the Voyager flyby, compared to a sub-observer latitude of $29^\circ$S during the Keck observations, such a mean is a valid approximation to the global average.   We used the exponential-integral method to calculate the disk-integrated spectrum \citep{89goody} and a spectral resolution of 0.2 $\mu$m ($R\approx100$).  The simulated radiance is almost indistinguishable from the calibrated Keck measurements, suggesting that there has been negligible change in the mean tropospheric temperature in the intervening 14 years.  As a final check, we performed a full optimal estimation retrieval to fit the 17.6-22.4 $\mu$m disk-averaged spectrum.  The best-fitting spectrum had a slightly lower radiance than the IRIS model (the dashed line in Fig. \ref{keckQ}), but given the size of the uncertainties on the Keck data (we estimate an accuracy no better than 10\%, given the scaling to produce the disk-integrated spectrum), such a shift is negligible.  Furthermore, the best-fitting temperature profile differed from the IRIS result by no more than 0.7 K at the tropopause, well within the formal retrieval uncertainties.  We therefore conclude that the Keck/LWS spectrum is consistent with the Voyager/IRIS temperature profiles, and that there has been no change in the globally-averaged tropospheric temperatures.

\section{Synthetic Images}
\label{images}

\begin{figure*}[tbp]
\centering
\includegraphics[width=13cm]{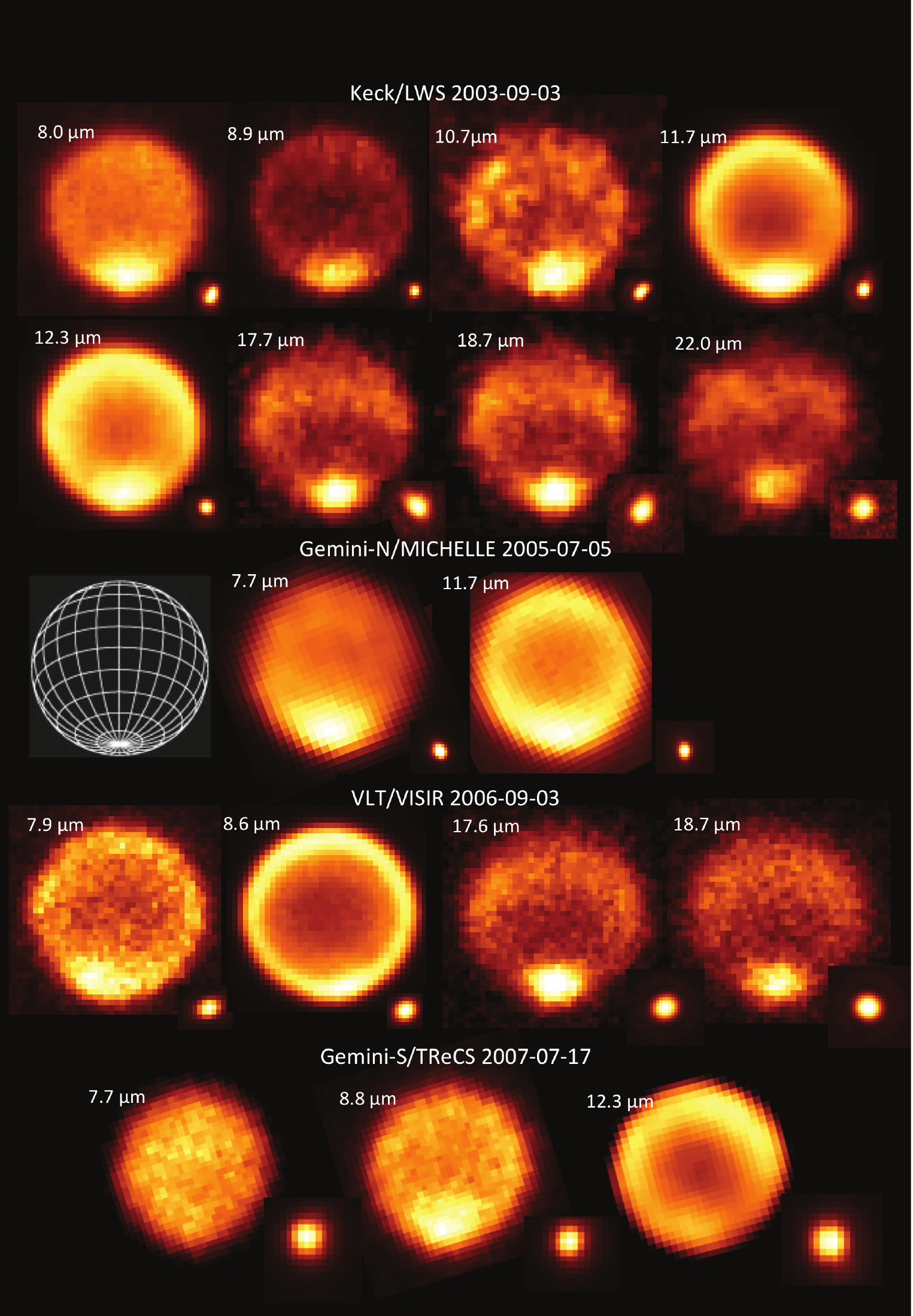}
\caption{Thermal-IR images of Neptune considered in this study.  Neptune has been oriented so that north is up, as shown by the guideline grid in the centre of the figure.  A representative stellar point spread function (PSF) obtained close to the Neptune image is given to the lower right of each image, showing how the PSF grows larger with wavelength.  Further details are given in Table \ref{imagetab}.  Offset polar hotspots, evident in the stratospheric images in 2006, were discussed by \citet{12orton}.}
\label{nepimages}
\end{figure*}

The majority of observations surrounding Neptune's solstice were images in a variety of filters sensitive to different altitudes in Neptune's atmosphere.  These images reveal the latitudinal variations in tropospheric and stratospheric emission, and were used to characterise Neptune's warm south polar vortex \citep{07orton, 07hammel, 12orton, 13depater}.  In this section, we use the meridional temperature structures derived in Sections \ref{stratos} and \ref{tropos} to generate synthetic images for direct comparison with the thermal-IR images between 2003 and 2007, shown in Fig. \ref{nepimages}.  Our aim is to search for evidence of variability in meridional temperatures between 1989 and 2007.

\subsection{Sources of Imaging Data}

Fig. \ref{refspx} compares a reference spectrum of Neptune (calculated at the native spectral resolution of Voyager/IRIS, using the $T(p)$ profiles derived in the previous section) to the positions of filters on a variety of mid-IR imagers at ground-based observatories.  Each of these imaging datasets have been previously published, our intention here is to analyse them using a consistent approach.  We have not deconvolved any of the images, so that the variability is not compromised by artefacts associated with the deconvolution process.


\textbf{Keck/LWS (2003):} \citet{13depater} present Neptune images obtained on September 5, 2003 (05:49-08:42 UT) using the Long Wavelength Spectrometer (LWS) at the W.M. Keck I telescope in Mauna Kea, Hawaii. The LWS detector is a 128 x 128 Boeing Si:As array, with a pixel size of 0.0847" \citep{93jones}. Given the large primary mirror diameter, this dataset represents the highest spatial resolution dataset ever obtained from Earth. The authors used a total of eight mid-infrared filters probing stratospheric hydrocarbon emission due to roto-vibrational transitions of methane (8.0 $\mu$m, 8.9 $\mu$m), ethylene (10.7 $\mu$m) and ethane (11.7 $\mu$m and 12.5 $\mu$m), in addition to molecular hydrogen emission due to collision-induced absorption in the troposphere (17.65 $\mu$m, 18.75 $\mu$m and 22 $\mu$m).   The reduction of the chop-nod differenced images to calibrated images \citep[using HD199345 as a standard calibrator,][]{99cohen} with a 10\% flux uncertainty is described by \citet{13depater}.  The images are shown as a montage in Fig. \ref{nepimages}.    

\textbf{Gemini-North/MICHELLE (2005):}  \citet{07hammel} obtained images at 11.7 and 7.7 $\mu$m on July 5, 2005 (10:43-14:16 UT) using the MICHELLE mid-infrared imager and spectrometer \citep{97glasse} on the Gemini-North 8-m telescope.  The MICHELLE $320\times240$ array consists of Raytheon Si:As detector, with a plate scale of 0.099"/pixel.  The on source exposure time was 120 seconds, and images were calibrated using standard star HD 199345 \citep{99cohen} for flux calibration.  For full details of the image processing techniques, the reader is referred to \citet{07hammel}.  

\textbf{VLT/VISIR (2006):}  Thermal images were obtained by ESO's Very Large Telescope (VLT) UT-3 (Melipal) mid-infrared camera/spectrometer \citep[VISIR,][]{04lagage} in September 2006, and presented by \citet{07orton}.  In this analysis we use an 18.7-$\mu$m image from 01:44-03:47 UT on September 2, 2006; and images at 17.6 $\mu$m (01:55-03:13 UT), 8.6 $\mu$m (06:53-07:40 UT) and 12.3 $\mu$m (03:16-03:36 UT) from September 3, 2006.  The data reduction and calibration are described in \citet{07orton}. Each image used a different standard star \citep{99cohen}.  Although \citet{07orton} have already provided a detailed comparison of the VISIR results to the Voyager/IRIS temperature retrievals, we repeat the analysis here to ensure consistency between all available datasets.

\textbf{Gemini-South/TReCS (2007):} A subset of Neptune images obtained with the Thermal-Region Camera Spectrograph on Gemini-South in 2007 \citep[TReCS,][]{05debuizer} were presented by \citet{12orton}.  TReCS uses a $320\times240$ Raytheon Si:As array with a 0.09"/pixel plate scale for narrow-band filtered imaging between 7-26 $\mu$m.  Gemini-South attempted to observe Neptune on multiple dates throughout 2007 (July 17, August 30, September 5, 10, 13 and 27), but was hampered by poor weather conditions, variable water vapour and high airmass observations.  The August 30 and September 13 observations focussed on the N-band spectroscopy analysed in Section \ref{stratos}.  Of all the dates, July 17 and September 10 provided the largest number of different filters.  However, inspection of the reduced images indicated that the atmospheric seeing was more favourable in July, so we select the following images for modelling:  Si-6 12.3 $\mu$m, 480 seconds of integration between 05:54-06:22 UT; Si-1 7.7 $\mu$m, 900 seconds of integration between 06:22-07:15 UT; Si-2 8.7 $\mu$m, 900 seconds of integration between 07:15-08:07 UT and Si-4 10.4 $\mu$m, 540 seconds of integration between 08:07-08:39 UT, although the latter was omitted from this study due to poor signal-to-noise and the possibility of variable clouds.   Flux calibration used the star HD 216032. The 2007 observations, like those in 2005, were mainly sensitive to stratospheric temperatures, so will be used to assess the validity of the uniform stratospheric temperatures derived in Section \ref{stratos}.

\begin{sidewaystable}[htp]  
\small
\caption{Sources of Neptune imaging 2003-2007, showing the heliocentric longitude ($L_s$) of the observation in degrees, with $L_s=270^\circ$ being summer solstice.}
\begin{tabular}{ccccccc}
\hline
\hline
Source & Primary & L$_s$ & Date & Times  &  Filters & Reference \\ 
& Diameter [m] &  & (UT) &  (\m) &  \\ 

\hline\\
Keck/LWS & 10.1 & 266.0 & 2003-09-05 & 05:49-08:42 UT & 8.0, 8.9, 10.7, 11.7,  & \citet{13depater} \\
 & & & & &12.5, 17.65, 18.75, 22.0 & \\
Gemini-N/Michelle & 8.1 & 270.0 & 2005-07-05 & 10:52-14:16 UT & 7.7, 11.7 & \citet{07hammel} \\
VLT/VISIR & 8.2 & 272.5 & 2006-09-02 & 01:44-03:47 UT & 18.7 & \citet{07orton} \\
VLT/VISIR & 8.2 & 272.5 & 2006-09-03 & 01:55-07:40 UT & 17.6, 8.6, 12.3 & \citet{07orton} \\
Gemini-S/TReCS & 8.1 & 274.4 & 2007-07-17 & 05:54-08:07 UT & 7.9, 8.6,12.3 & \citet{12orton} \\
\hline\\
\end{tabular}

\label{imagetab}
\end{sidewaystable}

\subsection{General Characteristics of Spatially-Resolved Imaging}

\begin{figure*}[tbp]
\centering
\includegraphics[height=17cm]{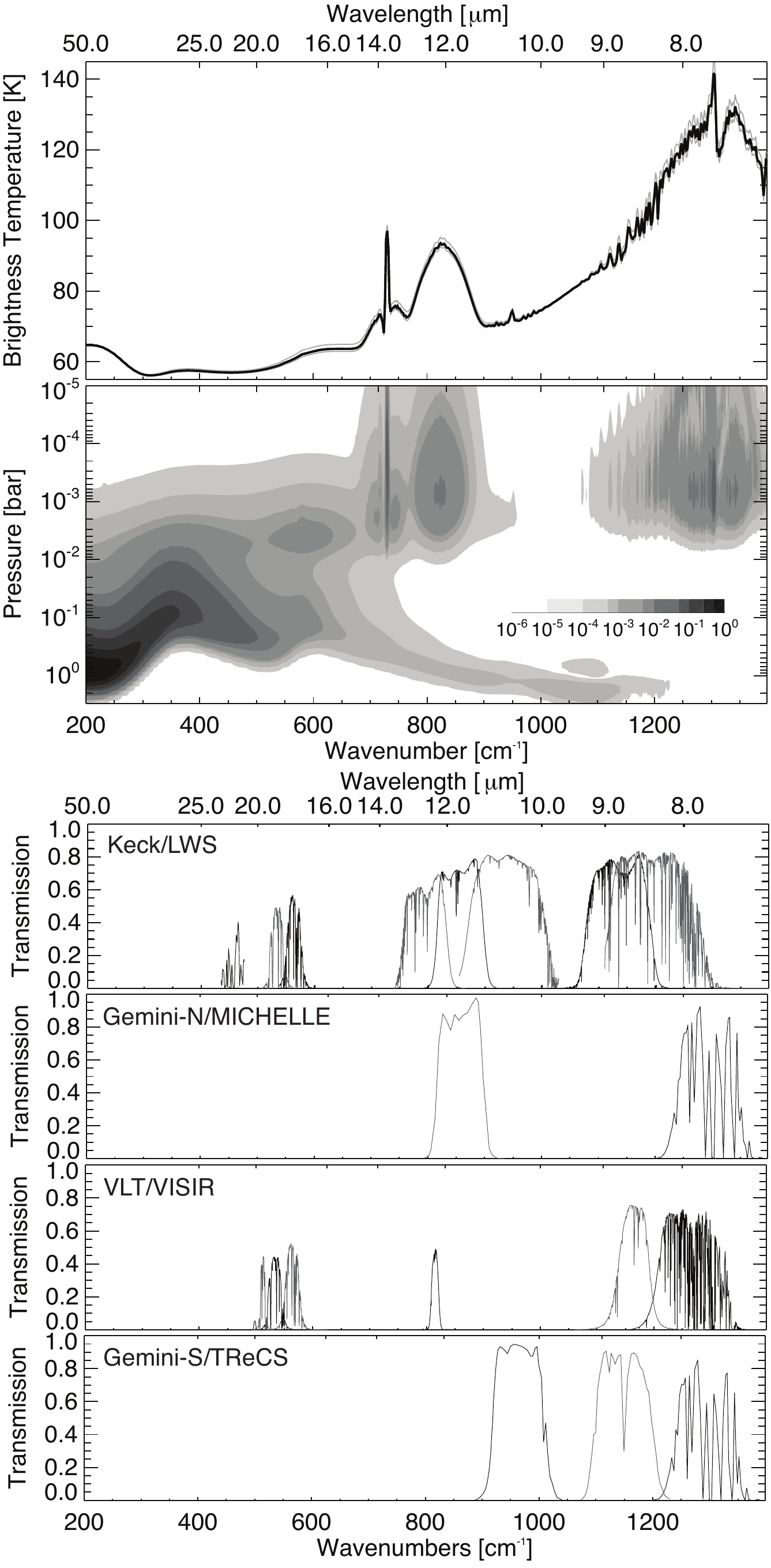}
\caption{Comparing the filters used in this study (bottom four panels) to a reference spectrum of Neptune (top panel) and contribution functions calculated for nadir viewing (central panel).  The reference spectrum is based on the mean $T(p)$ from all solstice measurements, the grey spectra either side of the solid line represent uncertainty based on the standard deviation of the $T(p)$ profiles.  The spectrum was calculated at the spectral resolution of Voyager/IRIS (4.3 cm$^{-1}$).  The scale in the contribution function plot is logarithmic so as to highlight the problems of multi-lobed functions are certain wavelengths, and the key is shown inset.  Adjacent filters are shown in different shades of grey to ease comparison, and are convolved with telluric transmission.}
\label{refspx}
\end{figure*}

A montage of the best examples of images from 2003-2007 is shown in Fig. \ref{nepimages}.  The appearance depends upon the placement of the filter with respect to Neptune's emission spectrum, shown in Fig. \ref{refspx}. In general, those images sensitive to the tropospheric collision induced continuum of H$_2$-H$_2$ and H$_2$-He indicate a warm equator and southern high latitudes, along with cool mid-latitudes, qualitatively consistent with Voyager/IRIS results in 1989 (Section \ref{tropos}).  Stratospheric images are dominated by centre-to-limb brightening, and typically appear as a bland disc with a bright limb.   The limb does not always appear uniform in brightness, a possible indication of zonal or meridional inhomogeneity, but such an effect can also be reproduced by an asymmetric point spread function (PSF).  At 11.7-12.3 $\mu$m, sensitive to emission from stratospheric ethane, the bright south pole merges with the ring of bright emission at the limb.  This is not the case at 7.7-8.9 $\mu$m, sensitive to emission from methane and its deuterated isotopologue (CH$_3$D), where the bright south pole dominates the images.  A bright, compact region of stratospheric emission was occasionally observed to be offset from the south pole in this dataset \citep[September 2, 2006 and September 21, 2007,][]{12orton}, but here we focus on the details of the meridional structure.

\begin{figure*}[tb]
\centering
\includegraphics[width=17cm]{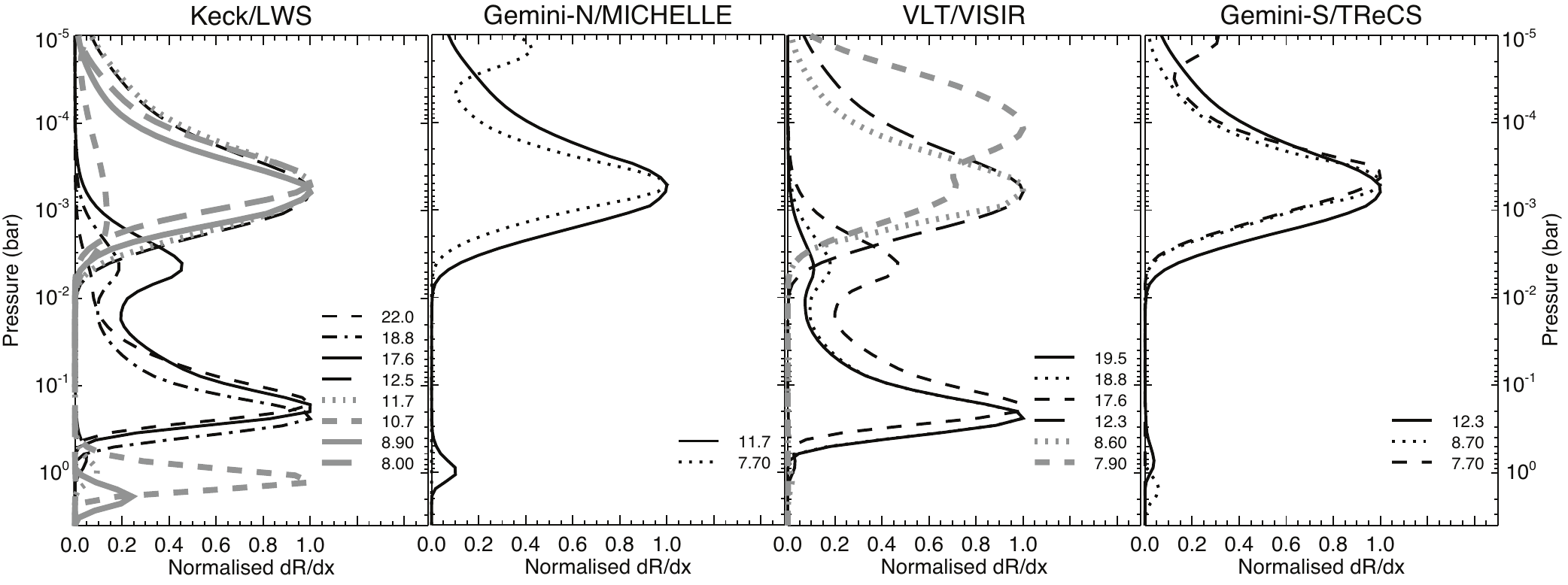}
\caption{Contribution functions calculated for each of the filters used in this study from the four telescopes.  Small differences in the filter transmissions, when convolved with the telluric transmission, can have substantial impacts on the range of altitude sensitivity.  These calculations are all performed for nadir viewing, and normalised to unity for ease of comparison.  The key to the central wavelengths, in microns, is inset within each panel.}
\label{contributions}
\end{figure*}

To understand why each filter reveals a slightly different morphology, we calculate contribution functions for each filter in Fig. \ref{contributions}.  The sources of line data described in Section \ref{model} were used to generate $k$-distributions convolved with (i) the transmission of the filter and (ii) the transmissivity of the atmosphere, following techniques described by \citet{09fletcher_imaging}.  For each gas, a $k$-table consists of cumulative distributions of absorption coefficients (over 20 ordinates for integrating the $k$-distribution, 20 temperatures from 40-300 K and 20 pressures spanning the expected range found in Neptune's atmosphere), one for each of the filters in a particular instrument.   We have calculated $k$-tables for four instruments:  Keck/LWS\footnote{http://www2.keck.hawaii.edu/inst/lws/filters.html}, Gemini-N/MICHELLE\footnote{http://www.gemini.edu/?q=node/10021}, Gemini-S/TReCS\footnote{http://www.gemini.edu/?q=node/10073} and VLT/VISIR\footnote{http://www.eso.org/sci/facilities/paranal/instruments/visir/inst/index.html}, and compare them in Fig. \ref{contributions}.  The Keck/LWS 22-$\mu$m filter used in 2003 was later replaced, but we were able to obtain the relevant filter transmission function via observatory staff.  It is worth noting that the 8 $\mu$m filter is particularly broad, covering the 7.8-8.9 $\mu$m passband (as measured with the LWS spectrometer at the operating temperature of 5 K), and is in fact centred at 8.4 $\mu$m.  

The most striking feature of Fig. \ref{contributions} is the multi-lobed nature of some of these contribution functions, introducing an inherent degeneracy into their interpretation.  The Q-band filters sense temperatures in the 100-300 mbar range, with filters centred near 18.7 $\mu$m sensing deeper in the atmosphere than those at 17.6 $\mu$m (close the the S(1) absorption of H$_2$) or 22 $\mu$m (close to the S(0) absorption of H$_2$).  However, each of these filters has some residual sensitivity to the 3-6 mbar region, with 17.6 $\mu$m being the most effected and 22 $\mu$m the least.     Filters in the centres of the strong stratospheric emissions (8 $\mu$m and 12.3 $\mu$m) are solely sensitive to stratospheric altitudes (0.3-0.7 mbar).  However, filters on the wings of these features (8.9, 10.7 and 11.7 $\mu$m near emission lines of methane, ethylene and ethane, respectively) have some sensitivity to the 0.3-0.7 mbar region, and residual sensitivity to tropospheric temperatures near 1 bar from the H$_2$ and He continuum.  

\begin{figure*}[tb]
\centering
\includegraphics[width=12cm]{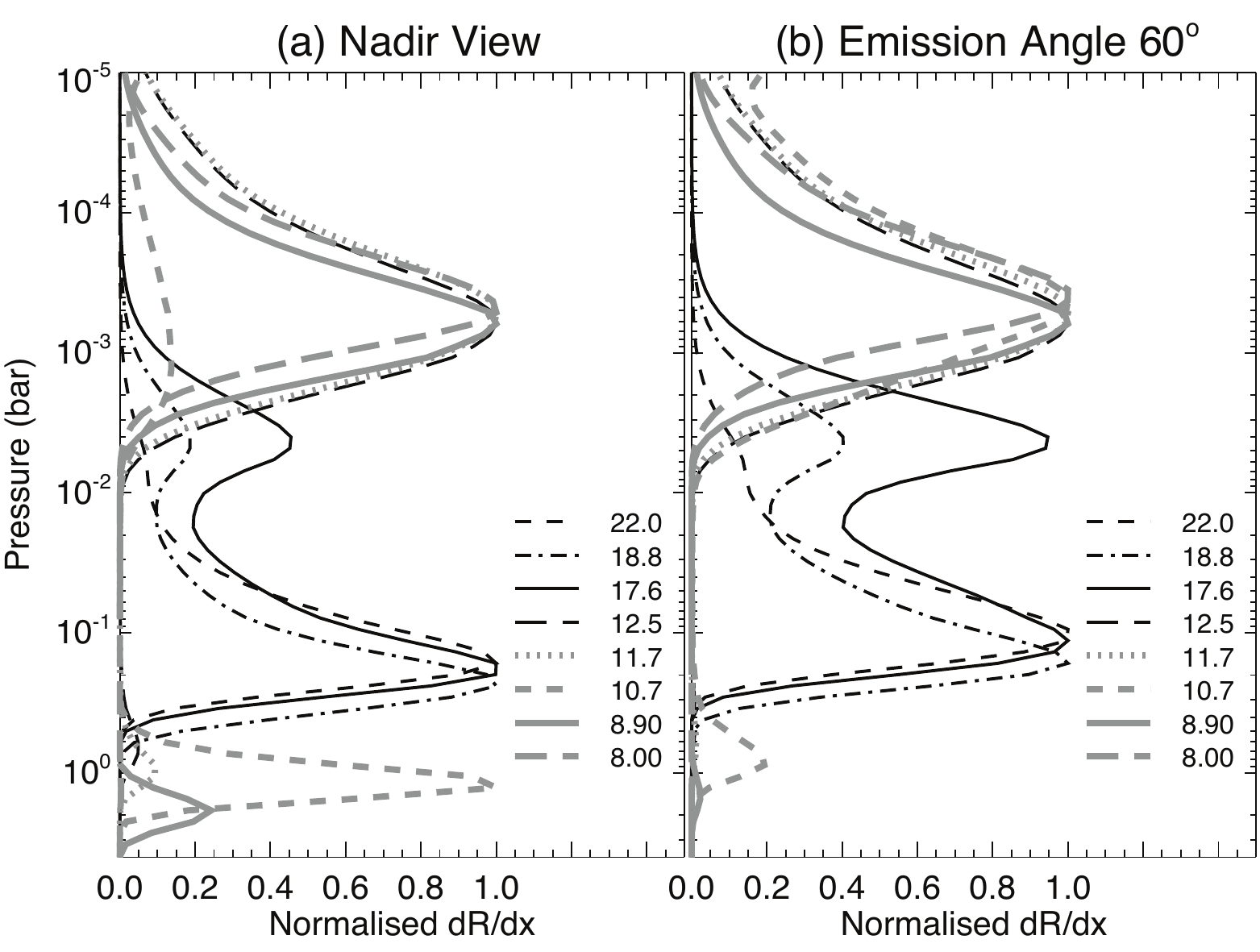}
\caption{Contribution functions for the Keck/LWS filters at two different emission angles (nadir and 60$^\circ$), showing how sensitivity changes from the nadir to the near-limb, including the increasing importance of the secondary stratospheric peak for 17-19 $\mu$m filters and the diminishing importance of the 1-bar sensitivity of 10.7-$\mu$m filters.}
\label{contrib_emm}
\end{figure*}

It is these multi-lobed contribution functions that pose the greatest difficulty in interpreting photometric imaging of Neptune.  Where hot spots are observed, for example at the south pole, there will remain uncertainties as to their precise altitude.  Furthermore, Neptune's high latitudes are observed at high emission angles, shifting contribution functions to even higher altitude (hence the limb brightening), as shown in Fig. \ref{contrib_emm} for the Keck/LWS filters.  For the Q-band filters, the 3-6 mbar residual peak becomes more significant for larger slant paths through the atmosphere. For N-band filters, the sensitivity to the 1-bar level wanes to insignificance.  Only the 22-$\mu$m filter retains a majority contribution from tropospheric levels.

\subsection{Generation of Synthetic Images}

\begin{figure*}[tb]
\centering
\includegraphics[width=13cm]{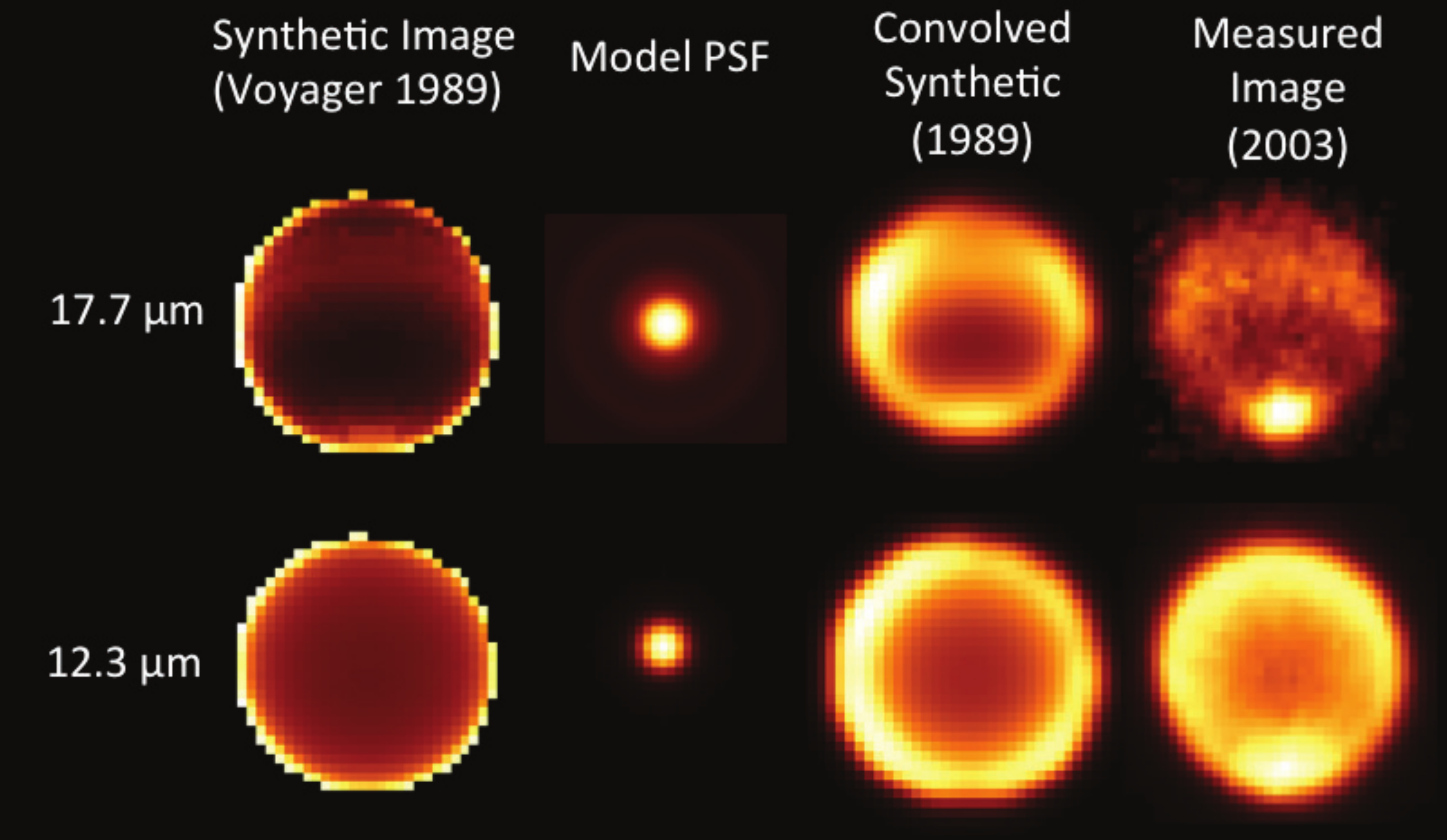}
\caption{An example of the procedure used to generate synthetic images for comparison with the data.  Combining Voyager/IRIS tropospheric temperatures with the solstice estimate of the stratospheric $T(p)$, we calculate synthetic radiances for all points on the disc and convolve with an idealised symmetric PSF.  Extraction of zonal mean spectra from both ensures an accurate comparison of conditions between 1989 and solstice.}
\label{synthetic}
\end{figure*}

For each of the four epochs, we determined the latitude, longitude and emission angle associated with each pixel in a representative Neptune image.  The plate scale of the instrument, sub-observer latitude and orientation of Neptune were all taken into account.  The Voyager-derived tropospheric temperature and para-hydrogen profile (Section \ref{tropos}) at the closest latitude for each pixel, along with a mean of the stratospheric temperatures and ethane abundance derived in Section \ref{stratos}, were used to calculate the flux for each pixel and each filter.  In essence, we are assuming that the stratospheric temperatures are uniform with latitude.  The radiances were reprojected to form an image of Neptune as it would be if there had been no changes since the 1989 Voyager measurements.  This image was convolved with an idealised point spread function (PSF), convolving an Airy function representing wavelength-dependent diffraction (with Bessel corrections to represent obscuration by the secondary mirror) and a Gaussian function representing seeing.  The parameters of this PSF were tuned to match the observed width of the rather noisy stellar images acquired close to the Neptune images, but have the benefit of being symmetric and noise-free.  The result is a synthetic image of Neptune as it would appear through a particular telescope and filters, using the Neptune atmospheric properties determined for 1989. An example of this process for a Q-band and an N-band image is shown in Fig. \ref{synthetic}.

\begin{figure*}[tbp]
\centering
\includegraphics[width=12cm]{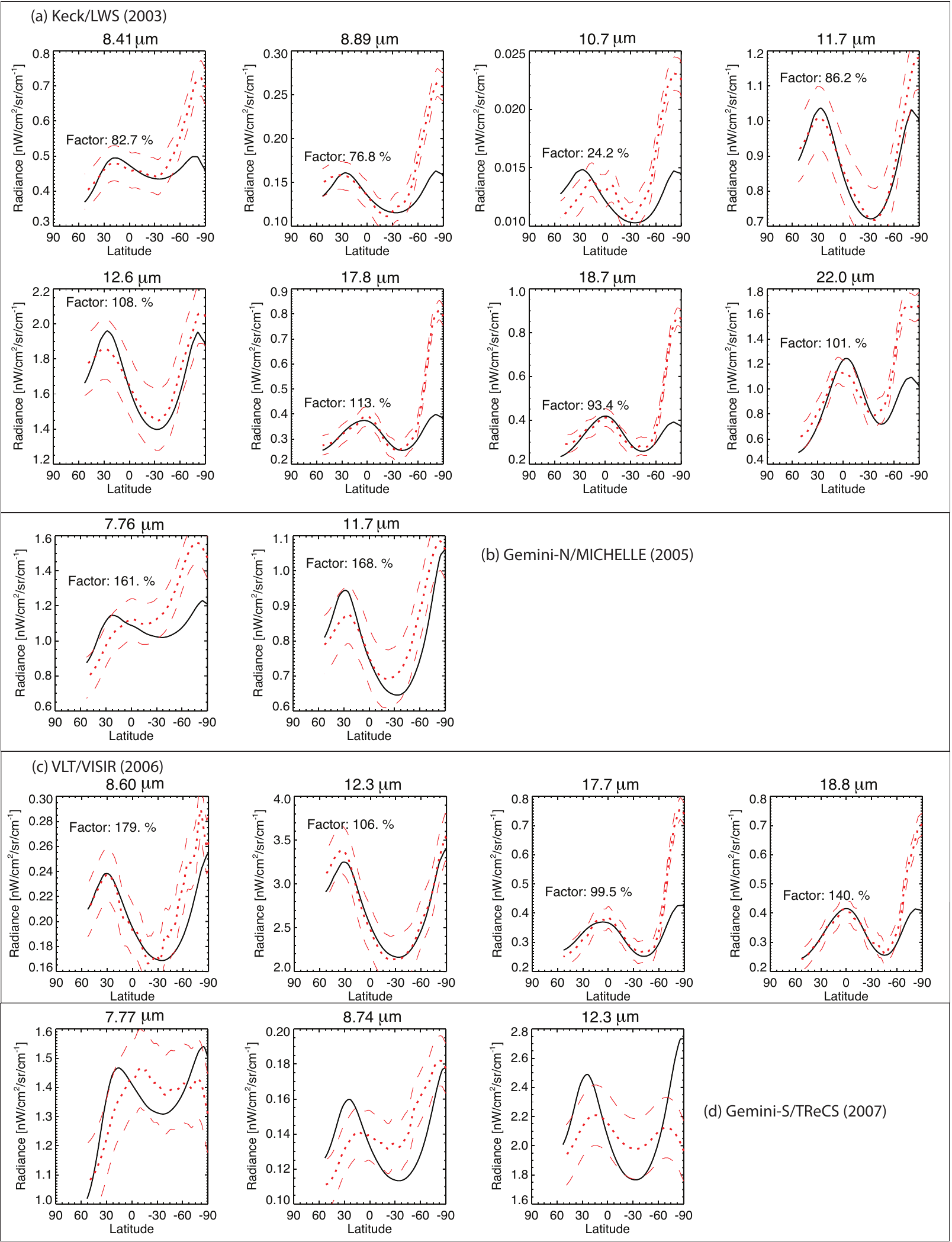}
\caption{Comparing zonal radiances along the central meridian for both the synthetic Voyager-era images (solid line) and the solstice-era measurements (dotted line).  The top panel is for Keck/LWS measurements (2003), followed by Gemini-N/MICHELLE measurements in 2005; VLT/VISIR measurements in 2006 and Gemini-S/TReCS measurements in 2007 in the bottom panel.   A 10\% uncertainty has been added to the relative variations observed in the solstice measurements (dashed lines).  The solstice measurements equatorward of $\pm45^\circ$ (i.e., avoiding the poles) were scaled by the amounts shown in each panel to permit a comparison of \textit{relative} contrasts in the zonal mean radiances.  A factor of 100\% implies that no scaling was necessary to match the measurements and model.  No absolute calibration was performed for the 2007 observations due to non-photometric conditions.  The scaling for the 2005 measurements was calculated based on the disc-averaged temperatures quoted by \citep{07hammel}. }
\label{zonalradiance}
\end{figure*}

We chose this forward modelling approach, rather than attempting to invert the photometric images to retrieve atmospheric temperatures, because of the multi-lobed contribution functions described earlier, and because of the small numbers of images obtained on each date.  The synthetic image and true image at each wavelength were then treated identically, reprojecting them to form a cylindrical map on a $1\times1$ degree grid.  Radiances were averaged within $\pm20^\circ$ of the central meridian, producing a near-nadir zonal mean for both the synthetic image and the true image, ensuring an accurate comparison of Voyager-era and solstice-era measurements in Fig. \ref{zonalradiance}.  Sources of error include (i) the difficulty in fitting the blurred planetary limb to calculate the geometry (latitudes, longitudes and emission angles) and (ii) uncertainty in the rotation of Neptune on the array due the the absence of a clear latitudinally-banded structure in the majority of the images.

\subsubsection{Radiometric Uncertainties}

However, it was soon discovered that some of the measurements had to be scaled to match the model.  The correspondence between model and data worked very well in the Q band, confirmed by the fact that the Keck/LWS Q-band spectrum and photometric images are consistent with one another \citep[as shown by][]{13depater}.  However, in the N band the model over predicted some measurements, and under predicted others.  Several tests were performed - regenerating $k$-distributions with different assumptions about telluric absorption; artificially narrowing and widening the manufacturer-provided filter profiles; and comparing forward model calibrations with (i) absorption coefficients pre-convolved with the filter functions (our preferred method) and (ii) convolving high-resolution spectra with the filter profiles post-calculation.  The latter test provided identical answers, validating our pre-convolution of the absorption spectra, whereas the other tests produced changes too small to explain the discrepancy between models and data.    Another possibility was variability on the planet itself from observation to observation, but inspection of images on consecutive nights \citep[e.g.,][]{12orton} did not reveal any large air masses that could be responsible for enhanced or reduced emission.  


We next considered the absolute calibration of the images, which ideally would use Cohen standard stars observed at almost the same time and airmass.  For Keck/LWS, it was noted that almost three hours elapsed between the Neptune observations and those of a standard star at the same wavelength, allowing for significant atmospheric changes during this interval.  Telluric extinction is particularly problematic towards the ends of the N band, near 8 and 12 $\mu$m, which is where our largest model-data discrepancies occur.  However, attempts to correct the Neptune images for reasonable assumptions of airmass and water vapour differences produced offsets that were too small to explain our discrepancies.  

Although the radiative transfer model for the spectra and images is internally consistent, and have reproduced the N and Q band spectra, we are unable to offer an explanation as to why we fail to reproduce the fluxes measured through \textit{some} of the filters.  We therefore scaled the measured data to ensure that the mean radiances equatorward of $\pm45^\circ$ were identical to the model, and quote the offset in Fig. \ref{zonalradiance}.  Although this prevents us from assessing absolute changes in Neptune's brightness \citep[e.g., the long-term study of][]{06hammel}, it does mean that relative variations can be studied.  Such an approach is analogous to the scaling of photometric imaging to match spacecraft data \citep{09fletcher_imaging}, and consistency checks with Spitzer data have been used by several previous authors in their studies of Neptune \citep[e.g.,][]{07orton,07hammel}.

\subsection{Changes from 1989 to Solstice}

The generation of synthetic Neptune images, viewed as though the Keck, Gemini and VLT were able to image the planet in 1989, should permit a direct comparison of results.  However, we have seen that radiometric uncertainties cause absolute offsets, and that the near-equatorial flyby geometry of Voyager 2 limited its ability to study the south pole.  Nevertheless, the comparison between the 1989 model and the solstice observations is qualitatively accurate:  it shows a troposphere with a warm equator, warm poles and cool mid-latitudes, and a stratosphere whose meridional variation is caused mostly by limb brightening (recall that the stratospheric temperatures are meridionally uniform in our model).  Considerable differences are seen in the ratio of the equatorial and polar radiances in the 17-22 $\mu$m filters (Fig. \ref{zonalradiance}) in 2003 (Keck) and 2006 (VLT), suggesting that the south polar region warmed considerably between the Voyager era and solstice.  Quantitatively, the 17.7, 18.7 and 22-$\mu$m brightness temperatures in 2003 are 3.5, 4.0 and 2.0 K warmer, respectively, than expected from the Voyager-era model.  In 2006, the 17.7 and 18.8 $\mu$m brightness temperatures of the south pole differ from expectations by 3.0 and 2.5 K, respectively \citep[consistent with values reported by][]{07orton}.  However, this result should be taken with caution, as Voyager/IRIS mostly sensed radiances long wards of 25 $\mu$m (a hint of south polar brightening is observed in images assembled from 28$\mu$m radiances, sensitive to the 50-100 mbar range).  If 17-25 $\mu$m images had been obtained in 1989 ($L_s=235.6^\circ$), would we have seen a warm polar vortex similar to solstice ($L_s=270.0^\circ$)?   We return to this question in the next section.


Moving to the stratosphere, the synthetic images at 7.7, 8.7, 11.7 and 12.3 $\mu$m all expect to see diminished flux in the northern hemisphere, poleward of $30^\circ$N, which is replicated in the solstice observations.  However, this is due to the convolution of the high northern latitudes with cold deep space by the point spread function, and should not be taken as an indication of cold temperatures in the northern winter hemisphere.  None of the synthetic models expect an equatorial maximum in the stratospheric emission \citep[unlike the Voyager/IRIS stratospheric profiles of the 13.7-$\mu$m acetylene peak by][]{91bezard}, and it is not observed in the data.  The coolest latitudes are centred on $30^\circ$S simply due to the observational geometry, for both the models and data, and the majority of the rise towards the south pole is due to limb brightening.  \citet{91bezard} found a stratospheric distribution resembling Neptune's tropospheric contrasts \citep[Fig. \ref{iris_Tp_fp} and ][]{89conrath, 98conrath} with cool mid-latitudes and an increasing radiance towards the south pole, requiring thermal contrasts of up to 15 K to explain.   This level of stratospheric variability is called into question by the latitudinally-uniform stratospheric temperatures measured by Keck (2003), Gemini-South (2007) and Gemini-N/TEXES spectroscopy \citep{11greathouse}, and by the images reported here.

The most intriguing variability is associated with the south polar stratosphere.  On some dates (for example, the VLT images in 2006), the contrast between the northern and southern limb (e.g., $30^\circ$N and $90^\circ$S in Fig. \ref{zonalradiance}) is adequately reproduced by the Voyager-era model.  On other dates, the southern limb is observed to be brighter than the northern limb (e.g., Keck observations in 2003 which show a well defined south polar vortex).  The brightness temperature difference between the southern and northern limb in 2003 amounts to no more than 1 K at 11.7 and 12.3 $\mu$m, increasing to 3 K for 8.0 and 8.9 $\mu$m (the latter probe slightly higher altitudes).  In the 2005 MICHELLE data, the 11.7-$\mu$m north-south contrast is 3.5 K, and there are hints of a 4.5 K difference in the rather noisy 7.7-$\mu$m image.  On the other hand, the 2006 VLT 12.3-$\mu$m brightness temperatures differ by $<0.5$ K, and the noisier 8.6-$\mu$m brightness temperatures differ by $\approx 1$ K.  Finally, the 2007 Gemini-S TReCS data suggest that the northern limb is \textit{brighter} at 12.3 $\mu$m by $\approx0.5$ K, but the 8.7-$\mu$m data still suggests a 1.5-K enhancement at the southern limb (the 7.7-$\mu$m data are so noisy as to be inconclusive).  These final observations were taken under poor weather conditions.  Given that Neptune's south polar region occupies only a small number of pixels in each image, and is convolved with nearby deep space by an amount dependent on the particular diffraction pattern of the telescope, the plate scale of the detector array and the quality of atmospheric seeing during the observation, we do not believe these year-to-year changes are significant.  We conclude that the majority of Neptune's south polar stratospheric emission is due to limb brightening, with the images showing the confinement of a \textit{slightly} warmer airmass (brightness temperature enhancements of 1-4 K) in the 0.1-1.0 mbar region over the pole.  

\subsection{Retrieved Temperatures from Stacked Images}

The synthetic image 'forward model' has identified tropospheric and stratospheric \textit{brightness temperature} contrasts at the pole in the solstice data, so it is reasonable to see if a full inversion of these data can be used to retrieve the \textit{physical temperature} contrasts.  Similar retrievals from ground-based imaging using the NEMESIS retrieval code \citep{08irwin} have been successful on Jupiter and Saturn \citep{09fletcher_imaging}, but the small size of Neptune, the radiometric uncertainties and the fact that the most interesting contrasts are viewed at very high airmass (e.g., the pole, where deep space radiances are convolved with the Neptune flux to dilute the intensity) presents a unique challenge.  We choose to focus on the 2003 Keck observations, which have the largest number of filters sounding troposphere and stratosphere, stacking the images to form a crude spectrum of eight points.  The LWS zonal radiances from Fig. \ref{zonalradiance} were scaled to match the 1989 model at low latitudes (equatorward of $\pm45^\circ$), and interpolated onto a $3^\circ$ latitude grid.  To facilitate comparison, the Voyager-era model in the same eight filters was handled in an identical manner.  In doing so, the problem of beam dilution towards the limb should be reduced when we take the difference.

Given the small number of data points ($m=8$) compared the to the number of levels in our profile ($n=100$ levels, although the use of strong vertical smoothing means that they are not independent) means we have a highly under constrained ($m<n$) and ill-conditioned retrieval problem.  The degree of vertical smoothing was made large to remove non-physical oscillations in the final profile.  The 17.6- and 18.8-$\mu$m filters have an increasingly important contribution from the stratosphere at the highest emission angles (Fig. \ref{contrib_emm}), so there is an uncertainty as to whether the bright polar emission is tropospheric or stratospheric.  The same is true of the 8-11 $\mu$m filters, which provide both 1-bar sensitivity and sensitivity to the stratosphere.  This degeneracy is partially broken by the observation at 22 $\mu$m, which has only weak sensitivity to the stratosphere (e.g., Fig. \ref{contrib_emm}), and suggests that the warming must also extend into the troposphere in the 100-200 mbar region.  At the highest emission angles, only the 22 $\mu$m filter can constrain the tropopause temperatures, all the remaining LWS filters have a degeneracy with the stratosphere in the 0.1-10 mbar region, so we cannot determine whether the warm polar vortex penetrates any deeper.

\subsubsection{Retrieval Tuning}

In an attempt to break these degeneracies, we performed a test to see whether a single Gaussian temperature perturbation, centred on the tropopause and extending upwards and downwards, could explain all the observations at the south pole.  A 4-K perturbation is needed to explain the 22-$\mu$m polar enhancement, but this was inconsistent with the 7-9 K required for the 17-19 $\mu$m filters and the 10-K needed for the 10.7-$\mu$m filter.  If we move the perturbation to the stratosphere at 5 mbar, we find that a $\approx 20$ K perturbation can reproduce the 17-19 $\mu$m enhancements observed in 2003, but this over-predicts the 12.3 $\mu$m emission by a large amount.  In conclusion, a perturbation at a single altitude cannot reproduce the Keck 2003 data, and the most likely scenario is that polar warming occurs from the stratosphere all the way down to the tropopause region.

\begin{figure*}[tb]
\centering
\includegraphics[width=14cm]{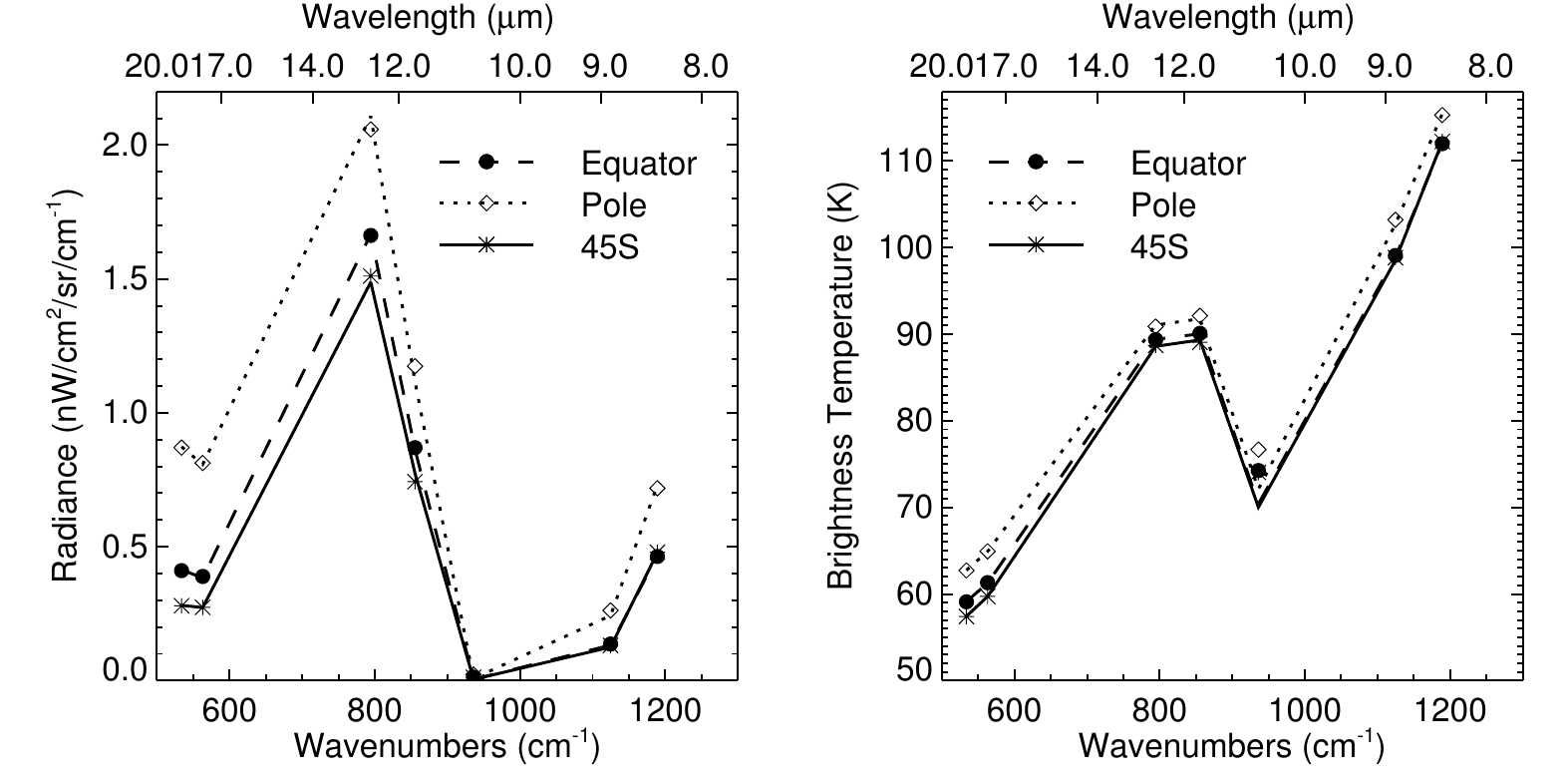}
\caption{Three examples of the fits to the 2003 Keck/LWS spectra formed from seven narrow-band images (i.e., excluding the 22 $\mu$m filter), represented as radiances on the left and brightness temperatures on the right.  Fits (lines) to the data (points) are shown for the warm equator and pole, and for the cool mid-latitudes.  The Q-band filters differ between these three locations, consistent with the tropospheric variability observed.  Methane emission is similar for the equator and mid-latitudes, consistent with the meridionally uniform stratospheric temperatures, but the emission is significantly enhanced at the pole.  }
\label{plotspx}
\end{figure*}

Despite numerous tests, we found no way to reproduce the radiances in all eight filters simultaneously (allowing $T(p)$ and ethane to vary as free parameters), particularly at high southern latitudes.  In searching for the problematic filters, we found the inconsistency between the zonal mean radiances in the 22 $\mu$m and 17-19 $\mu$m filters to be the source of the problem (e.g., Fig. \ref{nepimages}), and the elimination of the 22-$\mu$m filter allowed a clean fit to the 7-point spectrum (shown in Fig. \ref{plotspx}).  We suggest that the longest-wavelength filter was problematic for three reasons: (i) it suffers the greatest beam dilution due to the largest diffraction-limited seeing; (ii) the 22-$\mu$m filter is considerably broader and more affected by telluric absorption than all the other; and (iii) the 22-$\mu$m image was obtained on a different date to the others in this set.  In any case, the contribution functions of the 17-19 $\mu$m and the 22 $\mu$m filters were so similar that they should have been sensing the same altitude ranges.   The differences in north-south contrasts observed in these three filters prevented the retrieval algorithm from converging onto a solution consistent with all the available data.

\begin{figure*}[tbp]
\centering
\includegraphics[width=16cm]{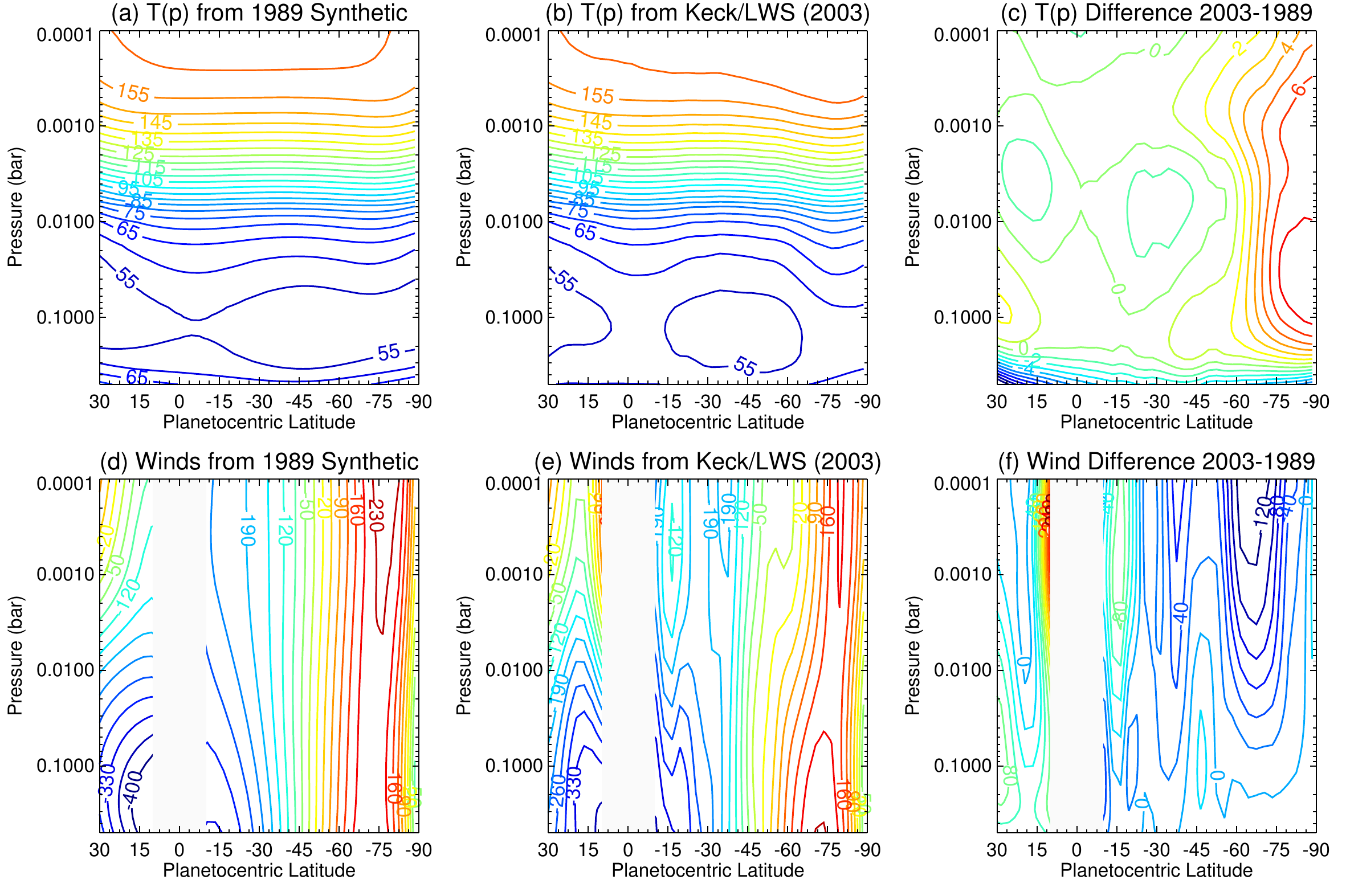}
\caption{Temperatures and winds retrieved from Keck/LWS imaging in 2003 (b, e) compared to retrievals from a synthetic filter set based on Voyager-era $T(p)$ profiles (a,d).  The differences in temperatures and winds are shown in panels (c) and (f).  Beam dillution with deep space pixels is noticeable at the furthest points from the sub-observer latitude ($28^\circ$ S), but as both the Voyager and Solstice-era images were handled identically, these should be removed by taking differences.  Latitudes within $10^\circ$ of the equator have been omitted in the lower row, as the geostrophic assumptions underlying the thermal windshear balance are inappropriate close to the equator.}
\label{imageret}
\end{figure*}

\subsubsection{Temperatures and Winds from 1989 to 2003}

Fig. \ref{imageret} shows the results of the 7-filter retrieval to the zonal mean radiances (i) synthesised based on Voyager/IRIS for 1989; and (ii) measured near solstice in 2003.  The first thing to note is the general trend to be cooler at the northern and southern edges of the domain - this is a spurious effect due to signal dilution at the increasingly high emission angles, which can be removed by taking differences.  Both the Voyager and solstice retrievals indicate a warm equator, cool mid-latitudes and warmer pole, qualitatively consistent with previous results \citep{89conrath}.  Subtracting the two results, we see differences $<1$ K between $\pm50^\circ$, smaller than our estimated uncertainty of 1.5-2.0 K.  The major difference is at the pole, where a maximum temperature enhancement of 7-8 K exists over the 10-100 mbar range, and an increase of 5-6 K exists throughout the 70-90$^\circ$ S region between 0.1 and 200 mbar.  Note that the contribution functions do not cover this full domain equally:  the profiles are interpolated between the three peaks of maximum sensitivity near the tropopause, 3-6 mbar and 0.1-1.0 mbar (e.g., Fig. \ref{contributions}).  This south polar temperature enhancement is therefore non-unique, but is the best-fitting solution reproducing the enhancements in the seven Keck/LWS filters.

\begin{figure}[tbp]
\centering
\includegraphics[width=5.5cm]{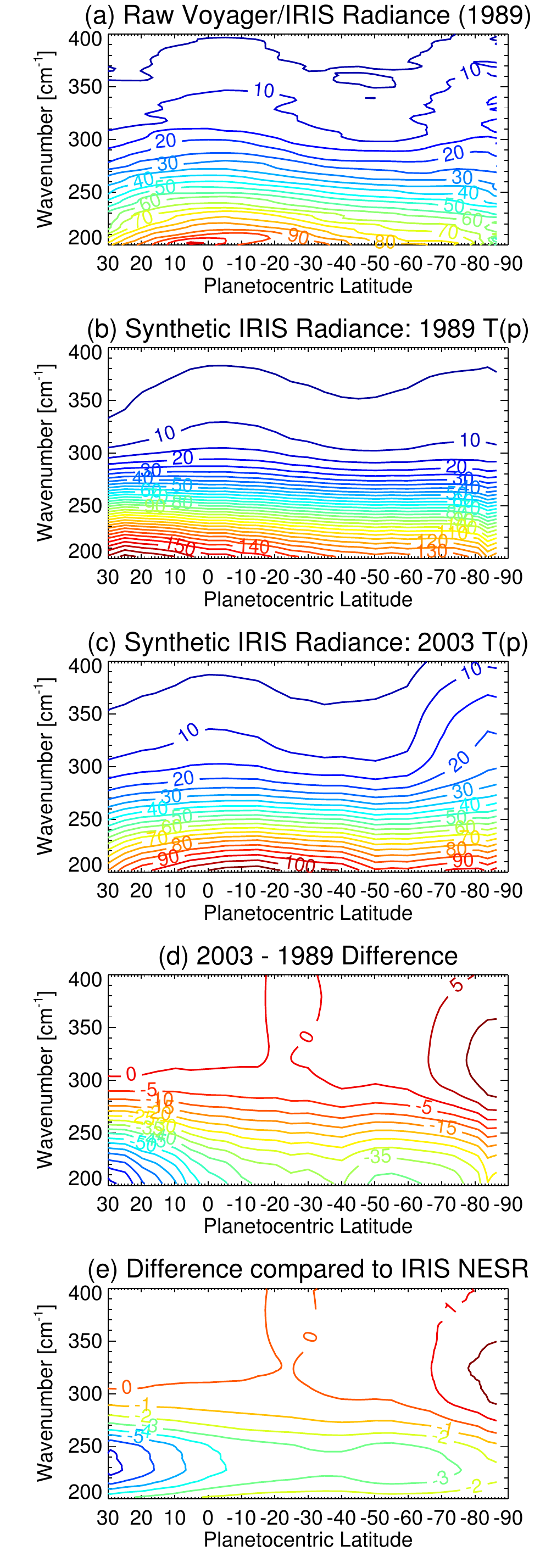}
\caption{Estimating the detectability of the warm polar vortex in the Voyager/IRIS measurements.  Panels (a)-(d) are given in units of radiance (nW/cm$^2$/sr/cm$^{-1}$) over the 200-400 cm$^{-1}$ region of maximal IRIS sensitivity.  Panel (a) gives the raw radiances from the outbound scanning maps, smoothed by a factor of three.  Panel (b) is a forward model based on the $T(p)$ structure retrieved from synthetic 1989 images, but using the Voyager encounter geometry.  Panel (c) is the same as (b), except for using the $T(p)$ retrieved from the 2003 Keck/LWS narrow-band imaging with the warmer polar vortex.  Panel (d) is the difference between (c) and (b), showing the expected significance of the warm polar vortex for IRIS spectra; and panel (c) reexpresses this in multiples of the single scan NESR (noise equivalent spectral radiance) for IRIS, indicating that the warm pole could have been visible in a single scan.}
\label{fmiris}
\end{figure}

\textit{Would such a large increase in south polar temperature have been observed in the 25-50 $\mu$m Voyager/IRIS observations?}  We took the two zonal temperature cross-sections (one for solstice, one for Voyager) and used them to forward model the expected 25-50 $\mu$m radiances based on the observing geometry of Voyager 2 (Fig. \ref{fmiris}(b-c)).  As expected, the warm pole has a larger influence at the shorter IRIS wavelengths, where the contribution function (Fig. \ref{refspx}) probes lower stratospheric altitudes, but the signal-to-noise in the Voyager/IRIS spectra (Fig. \ref{fmiris}(a)) was poor.  However, re-expressing the signal difference (Fig. \ref{fmiris}(d)) in terms of the IRIS Noise Equivalent Spectral Resolution (NESR\footnote{http://pds-rings.seti.org/vol/VG\_2001/NEPTUNE/CALIB/}, Fig. \ref{fmiris}(e)) for a single spectrum, we conclude that a warm south polar vortex would have been detected by Voyager in a single IRIS scan (with a signal-to-noise ratio exceeding two) if it had been present in 1989.  A substantial caveat to this is that the retrieved $T(p)$ structure is based on only seven filters with problematic beam dilution effects at high southern latitudes.  

If the differences between the Voyager and solstice zonal temperatures are real, then this has implications for the zonal windshear at high southern latitudes as the polar vortex developed.   Vertical wind shears are calculated by assuming geostrophic balance between the meridional pressure gradients and Coriolis forces.  In log-pressure coordinates, the thermal windshear equation for the zonal ($u$) direction is:
\begin{equation}
f\pderiv{u}{\ln \left( p \right)} = \frac{R}{a} \pderiv{T}{\psi} = R\pderiv{T}{y}\\
\end{equation}
where $T$ is the temperature in Kelvin; $p$ is the pressure in bar; $f$ is the Coriolis parameter $f=2\Omega \sin(\psi)$ (where $\Omega$ is the planetary angular velocity, $\psi$ is the latitude); $a$ is the mean planetary radius and $R$ is the molar gas constant divided by the mean molar weight of Neptune's atmosphere.  We estimate $u$ at the cloud-tops using a polynomial fit to the Voyager 2 wind data \citep{01sromovsky}, and integrate this with height using the thermal wind balance equation.  The warm polar stratosphere implies $\pderiv{T}{y}<0$ across the 70$^\circ$S boundary, so $\pderiv{u}{z}<0$ and the prograde jet peak encircling the pole at 70$^\circ$S decreases in strength with altitude.  The warm equator means that $\pderiv{u}{z}>0$, so the broad retrograde jet \citep[e.g.,][]{01sromovsky} decreases velocity with altitude.    In both cases, the dominant trend is for the zonal winds to decrease in strength with height due to poorly understood drag processes.  Between Voyager and Solstice, the polar jet at 70$^\circ$S has become more retrograde ($\pderiv{u}{y}$ has become more negative, meaning that the relative vorticity has become more positive, i.e., cyclonic), consistent with the onset of a cyclonic polar vortex in the stratosphere.  Polar cyclones are typically associated with subsidence and adiabatic warming, drying and cloud clearing.

Retrievals of temperature and winds from narrow-band imaging are extremely challenging because of the broad and multi-lobed appearance of the contribution functions, coupled with the high viewing angles for Neptune's poles.  Unfortunately, the imaging sets after 2003 have not had sufficient numbers of filters to permit a well-constrained retrieval, and the Voyager/IRIS spectra did not cover the N- and Q-band wavelengths necessary too confirm the absence of a distinct stratospheric vortex in 1989.  Nevertheless, a comparison of the retrieved solstice temperatures with the Voyager spectra do suggest that Neptune's warm stratospheric vortex formed (or at least strengthened) in the 14 years since the Voyager flyby.

\section{Discussion and Conclusions}
\label{discuss}

This study analysed four years of solstice imaging and spectroscopy of Neptune's thermal emission using a single, consistent radiative transfer and retrieval scheme to search for seasonal evolution since the Voyager-2 observations.  N-band (8-14 $\mu$m) disc-integrated and spatially-resolved spectra from 2003, 2005 and 2007 were used to determine the spatial variation of stratospheric temperatures and ethane in Section \ref{stratos}.  These properties were used as an \textit{a priori} to update the mapping of tropospheric temperatures and para-hydrogen distributions from Voyager/IRIS in 1989 in Section \ref{tropos}, and to confirm negligible changes to the disc-averaged 17-24 $\mu$m spectrum between 1989 and 2003 Keck observations.  Finally, Section \ref{images} used the zonal mean temperatures to forward model the expected radiances in narrow band imaging between 2003 and 2007; and the stacked set of seven Keck/LWS images was used to retrieve a solstice zonal temperature cross-section for comparison with Voyager/IRIS.  These images and spectra allow us to draw the following conclusions:

\begin{enumerate}
\item \textbf{Global Stratospheric Temperatures:  } Using Herschel estimates of Neptune's stratospheric methane abundance \citep{10lellouch}, the stratospheric temperatures tend to a quasi-isothermal structure for $p<0.1$ mbar (required to explain the absence of limb brightening in methane emission images) with temperatures ranging from 158.4 K (AKARI/IRC, 2007) to 164 K (Gemini-N/MICHELLE, 2005).  These 0.1-mbar temperatures show no consistent trend with time between 2003 and 2007, and differences may be solely due to the differing uncertainties on the data from each instrument.  A temperature gradient must exist in the lower stratosphere (1-10 mbar) to explain observations of limb brightening in H$_2$ S(1) emission \citep{11greathouse}.  We conclude that the global-mean stratospheric temperature is invariant during this period.  The overall consistency in $T(p)$ structure between 2003 and 2007 is remarkable given that the 8-9 $\mu$m region of the spectrum is plagued by telluric contamination, and the data do not cover the Q-branch of the CH$_4$ emission at 7.7 $\mu$m.  Stratospheric temporal variability, if present, is $<\pm5$ K at 1 mbar and $<\pm3$ K at 0.1 mbar.  Our derived $T(p)$ is consistent with the one used by \citet{05moses_jup} for photochemical modelling of Neptune; but 8-10 K warmer at 1-mbar than the $T(p)$ derived from ISO/SWS \citep{99bezard} and from Gemini-N/TEXES \citep{11greathouse}, although this discrepancy may arise from the absence of H$_2$ emission constraint in our $T(p)$ retrievals.  
\item \textbf{Ethane Emission Variability: } Despite the overall consistency, small changes in the $T(p)$ have substantial consequences for the retrieved ethane abundance.  Comparing ethane abundance retrievals with a fixed $T(p)$ over the four year period, we find that Keck/LWS (2003); Gemini-N/MICHELLE (2005) and AKARI/IRC (2007) provide consistent abundances, and hints of a decreasing mole fraction from 2003 to 2007.  However, Gemini-S/TReCS retrievals suggest a spike in the ethane abundance in the latter half of 2007 (i.e., the contrast between the ethane band at 12.3 $\mu$m and the methane band at 8.0 $\mu$m increased substantially).  Large abundance increases are considered unlikely, but significant temporal variability on ethane emission, apparently divorced from variability in methane emission, have been previously reported \citep{06hammel}.  It is possible that water ice cirrus clouds could be the cause, which would corrupt radiometric calibration near 12 $\mu$m but not near 8 $\mu$m.  Alternatively, we suggest that temperature variations in the upper troposphere and lower stratosphere (e.g., due to upward-propagating waves), undetected in methane emission, are leading to ethane emission variability over time.  The causes of these perturbations remain a mystery that will require a long-baseline of well-calibrated stratospheric spectroscopy to resolve.  Simultaneous observations of H$_2$ S(1) emission, along with methane and ethane, could help resolve this problem.
\item \textbf{Latitudinally-Uniform Stratosphere: } With the exception of the south polar vortex, the 2003 and 2007 N-band spectra reveal that Neptune's stratospheric temperatures are latitudinally uniform \citep[consistent with the 2007 Gemini-N/TEXES observations of][]{11greathouse}, and do not show the pronounced seasonal asymmetries observed on Saturn \citep{07fletcher_temp}.  This is based on the assumption of a latitudinally-uniform CH$_4$ distribution, which may be inaccurate if methane is being preferentially injected due to mid-latitude upwelling and convective overshooting \citep[e.g.,][]{98conrath}; equatorial upwelling \citep{11karkoschka_ch4}; or leakage through a warm tropopause at high latitudes \citep{07orton}, although the latter may be unlikely if the south polar vortex is a region of strong subsidence.  The latitudinally-uniform stratospheric emission is surprising when compared to Saturn's asymmetric emission \citep{07fletcher_temp} or radiative climate models \citep{11greathouse}.  However, as discussed by \citet{11greathouse}, Neptune would need to have depleted southern-hemisphere methane (e.g., due to polar subsidence) and enhanced northern-hemisphere methane in order to exhibit a seasonal asymmetry in stratospheric temperature matching the radiative models.  A depletion of south polar methane would imply that the vortex was even hotter than measured here, in order to produce the striking emission seen in our stratospheric images.  This mystery can only be resolved by independently constraining Neptune's stratospheric temperatures and methane distributions.  Finally, a cool stratospheric region between 40-50$^\circ$S in 2003 is reminiscent of the latitudinal profile of hydrocarbon emission measured by Voyager in 1989 \citep{91bezard}, but with a significantly smaller magnitude, and this was not reproduced in the 2007 Gemini data.  A latitudinally-uniform ethane distribution is consistent with the data, although the Keck 2003 data prefer a slight equator-to-pole decrease by a factor of $1.2\pm0.1$ that might support the theory of a depletion in stratospheric methane (the parent molecule) from equator to pole.  
\item \textbf{Tropospheric Invariance: }  At equatorial and mid-latitudes, both the comparison of synthetic and observed images, in addition to the retrievals of Keck (2003) and Voyager (1989) images and spectra, have suggested that the zonal mean temperatures in the troposphere did not change between 1989 and solstice.  The globally-averaged Q-band spectrum from Keck (2003) was consistent with that expected from Voyager measurements.  The overall contrast between the warm equator and cool mid-latitudes has remained unchanged, despite apparent changes in the activity levels of mid-latitude convective events.    
\item \textbf{Para-Hydrogen: }  Our re-analysis of Voyager/IRIS spectra using updated collision-induced H$_2$-H$_2$ absorption has revealed a different distribution of para-H$_2$ disequilibrium (a tracer for vertical motions) than that shown by \citet{98conrath}, presumably due to different assumptions about the collision-induced absorption coefficients.  The new results suggest para-H$_2$ disequilibrium is symmetric about the equator, with super-equilibrium conditions at the equator ($\pm20^\circ$) and at high southern latitudes, and sub-equilibrium conditions at mid-latitudes in both hemispheres.  This is consistent with a meridional circulation, with cold air rising at mid-latitudes and subsiding at both the poles and the equator.  Measurements of the para-H$_2$ distribution near solstice was not possible and should be a target for future observations.
\item \textbf{South Polar Vortex:  }  Neptune's most significant atmospheric changes have occurred at high southern latitudes.  Comparison of solstice images with synthetic Voyager-era images showed the confinement of a warm airmass over the south pole, with enhanced brightness temperatures of 1-4 K above that expected from limb brightening alone.  A retrieval from the Keck (2003) images revealed a polar enhancement of 7-8 K over the 10-100 mbar range, and an increase of 5-6 K exists throughout the 70-90$^\circ$S region between 0.1 and 200 mbar.  Forward modelling suggested that such a polar enhancement would have been detected by Voyager/IRIS in a single scan, despite its longer wavelengths, if the vortex had been present in 1989.  We conclude that the polar vortex, a warm airmass extending from at least 0.1-mbar down to the tropopause, has increased in temperature since 1989, coinciding with a decrease in strength of its prograde peripheral jet (i.e., increasingly cyclonic).
\end{enumerate}

The small angular size of Neptune, coupled with low atmospheric temperatures, presents a considerable challenge for thermal and compositional mapping, at the limits of 8-10-m class observatories available today.  The observed variability in ethane emission remains an outstanding problem, with uncertainties as to whether Neptunian temperature variations or terrestrial observing conditions are causing the changes.  Progress requires long-baseline, well-calibrated (preferably from space) spectral monitoring of Neptune over multiple years.  Higher spectral resolutions across a broad wavelength range will also be invaluable to distinguish changes in temperature from abundance variations, and an assessment of the stratospheric variability of methane is essential to reconcile temperature contrasts with seasonal radiative models \citep[e.g.,][]{11greathouse}.  Nevertheless, this combination of imaging and spectroscopy for the five years surrounding Neptune's solstice has revealed new insights into the slow seasonal changes occurring on our solar systems' most distant planet.




\section*{Acknowledgments}

Fletcher was supported during this research by a Royal Society Research Fellowship at the University of Oxford.  Some of the data presented here were obtained at the W.M. Keck Observatory, which is operated as a scientific partnership among the California Institute of Technology, the University of California and the National Aeronautics and Space Administration.  The Observatory was made possible by the generous financial support of the W.M. Keck Foundation.  This investigation was partially based on VLT/VISIR observations acquired at the Paranal UT3/Melipal Observatory under ID 077.C-0571; and on Gemini MICHELLE and TReCS observations acquired under ID GN-2005A-DD-10 and ID GS-2007B-Q-47 at the Gemini Observatory, which is operated by AURA, Inc., under an NSF cooperative agreement on behalf of the Gemini partnership.  

This work has been supported in part by the National Science Foundation Science and Technology Center for Adaptive Optics, managed by the University of California at Santa Cruz under cooperative agreement No. AST 9876783, and by NSF Grant AST-0908575 to UC Berkeley.  Orton was supported by grants fro NASA to the Jet Propulsion Laboratory, California Institute of Technology.  The UK authors acknowledge the support of the Science and Technology Facilities Council (STFC).

We thank T. Greathouse for providing the Neptune $T(p)$ structure derived for the October 2007 TEXES observations, and for interesting discussions about the nature of these results.  We also thank R. Campbell at Keck user support for providing details of the historical filters on Keck/LWS; and M. Gustafsson for providing collision-induced absorption coefficients for a range of para-H$_2$ fractions pertinent to Neptune's atmosphere.  Finally, we thank two anonymous reviewers for helping to improve the quality of this manuscript.  The UK authors acknowledge the support of the Science and Technology Facilities Council (STFC). 



\bibliographystyle{harvard}

\bibliography{../../../references_master}









\renewcommand{\baselinestretch}{1}
\onecolumn

\small





\label{lastfig}
\label{lastpage}
\end{document}